\documentstyle[prl,epsf,aps]{revtex}
  
\newtheorem{theorem}{Theorem}[section]

\newtheorem{remark}{Remark} 
 
\newtheorem{lemma}[theorem]{Lemma}
\newtheorem{proposition}{Proposition}
\newtheorem{corollary}{Corollary} 
\newtheorem{conjecture}{Conjecture}
\newtheorem{assumption}{Assumption}

\newcommand{\bd}{
\begin{document}} 
\newcommand{\ed}{\end{document}}
\newcommand{\beq}{\begin{equation}} 
\newcommand{\eeq}{\end{equation}}
\newcommand{\bef}{\begin{figure}} 
\newcommand{\enf}{\end{figure}}
\newcommand{\bea}{\begin{eqnarray}} 
\newcommand{\eea}{\end{eqnarray}}
\newcommand{\bth}{\begin{theorem}} 
\newcommand{\eth}{\end{theorem}}
\newcommand{\bhyp}{\begin{assumption}}
\newcommand{\ehyp}{\end{assumption}}
\newcommand{\bp}{\begin{proposition}}
\newcommand{\ep}{\end{proposition}}
\newcommand{\bco}{\begin{corollary}}
\newcommand{\eco}{\end{corollary}}
\newcommand{\bconj}{\begin{conjecture}}
\newcommand{\econj}{\end{conjecture}}
\newcommand{\ble}{\begin{lemma}} 
\newcommand{\ele}{\end{lemma}}
\newcommand{\bR}{\begin{remark}} 
\newcommand{\eR}{\end{remark}}
\newcommand{\bc}{\begin{center}} 
\newcommand{\ec}{\end{center}}
\newcommand{\ben}{\begin{enumerate}}
\newcommand{\een}{\end{enumerate}}
\newcommand{\bit}{\begin{itemize}} 
\newcommand{\eit}{\end{itemize}}
\newcommand{\su}{\section} 
\newcommand{\ssu}{\subsection}
\newcommand{\sssu}{\subsubsection} 
\newcommand{\nid}{\noindent}
\newcommand{\nnb}{\nonumber}

\newcommand\bbbr{{\sf I\!R}} 
\newcommand\bbbc{{\sf I\!\!C}}
\newcommand\bbbn{{\sf I\!N}} 
\newcommand\bbbh{{\sf I\!H}}
\newcommand\bbbz{{\sf Z\!\!Z}}

\newcommand\cA{{\cal A}} 
\newcommand\cB{{\cal B}} 
\newcommand\cC{{\cal C}} 
\newcommand\cD{{\cal D}} 
\newcommand\cE{{\cal E}}
\newcommand\cF{{\cal F}} 
\newcommand\cG{{\cal G}} 
\newcommand\cH{{\cal H}} 
\newcommand\cI{{\cal I}} 
\newcommand\cK{{\cal K}}
\newcommand\cL{{\cal L}} 
\newcommand\cM{{\cal M}} 
\newcommand\cN{{\cal N}} 
\newcommand\cP{{\cal P}} 
\newcommand\tcP{\tilde{\cal P}} 
\newcommand\cQ{{\cal Q}}
\newcommand\cR{{\cal R}} 
\newcommand\cS{{\cal S}} 
\newcommand\cW{{\cal W}} 
\newcommand\cU{{\cal U}} 
\newcommand\cV{{\cal V}}
\newcommand\cT{{\cal T}} 
\newcommand\cX{{\cal X}} 
\newcommand\cZ{{\cal Z}} 
\newcommand\cl{{\cal l}} 
\newcommand\cn{{\cal n}}

\newcommand\hX{\hat{\bX}} 
\newcommand\hY{\hat{\bY}} 
\newcommand\hF{\hat{\bF}}
\newcommand\hm{\hat{\mu}}
\newcommand\bE{{\bf E}} 
\newcommand\bF{{\bf F}} 
\newcommand\bJ{{\bf J}}
\newcommand\bX{{\bf X}} 
\newcommand\bY{{\bf Y}} 
\newcommand\bU{{\bf U}}
\newcommand\bZ{{\bf Z}} 
\newcommand\ba{{\bf a}} 
\newcommand\bb{{\bf b}} 
\newcommand\be{{\bf e}}
\newcommand\bk{{\bf k}}
\newcommand\bh{{\bf h}} 
\newcommand\bi{{\bf i}} 
\newcommand\bl{{\bf l}} 
\newcommand\bn{{\bf n}} 
\newcommand\bq{{\bf q}} 
\newcommand\bv{{\bf v}}
\newcommand\bx{{\bf x}} 
\newcommand\by{{\bf y}} 
\newcommand\bz{{\bf z}} 
\newcommand\bw{{\bf w}}
\newcommand\btX{{\bf \tilde{X}}} 

\newcommand\ti{\tilde{\i}} 
\newcommand\th{\tilde{h}} 
\newcommand\tbeta{\tilde{\beta}} 

\newcommand\bcm{\bar{\cM}} \newcommand\cm{\cM}

\newcommand{\deq}{\stackrel {\rm def}{=}}

\newcommand{\sep}{\; \,} \newcommand{\D}{\displaystyle}
\newcommand{\T}{\textstyle} \newcommand{\etc}{etc $\dots $ }
\newcommand{\etal}{etc $\dots$} \newcommand\dL{^\partial \Lambda}

\def\Appendix{\section*{APPENDIX}}

\baselineskip18pt

\bd
\draft

\baselineskip18pt

\title {Anomalous scaling and Lee-Yang zeroes in Self-Organized
Criticality.}

\author {
B. Cessac,
 J.L. Meunier,\thanks{Institut Non Lin\'eaire de Nice, 1361 Route des
Lucioles, 06500 Valbonne, France}
}

\maketitle

\begin{abstract}
We show that the generating functions
of probability distributions in SOC models exhibit a
Lee-Yang phenomenon \cite{LeeYang}. 
Namely, their zeroes pinch the real axis at $z=1$,
as the system size goes to infinity. This
establish a new link between the classical theory of critical phenomena
and SOC. A scaling theory of the Lee-Yang
zeroes is proposed in this setting.    
\end{abstract}


\bigskip

 In 1988, Bak, Tang and Wiesenfeld 
\cite{BTW}  proposed for the first time a mechanism in which 
a  dynamical system reaches ``spontaneously'' a stationary
state with some features reminiscent of a critical state.
More precisely, by its only
internal reorganization in reaction to (stationary)
 external perturbations, a system organizes into a state
with scale invariance and power law statistics. This effect, called
 Self-Organized Criticality (SOC), was quite
unexpected, since usually, the critical state of a thermodynamic
system needs a fine tuning of some control parameter (temperature, magnetic
field, etc...)
which is at first sight absent from the model introduced by BTW 
and from the many variants proposed later \cite{Bak1,Jensen}.
Furthermore, the stationary regime corresponds to a non-equilibrium
 situation where the (stationary) flux of external perturbation  is dissipated
in the bulk or at the boundaries, generating  a constant flux
 through the system. As a consequence, one generally believes that the usual
equilibrium statistical mechanics treatments using the concepts
of Gibbs measure, free energy, etc ...
 cannot be applied for the analysis of SOC systems. 

On the other hand, it is also believed that concepts like universality classes,
 critical exponents, order
parameter, \etc, borrowed from the  equilibrium statistical mechanics of
phase transitions are still relevant in SOC.
 Actually, the identification of universality classes
is one of the main goal in the SOC literature.
 However, since these concepts are not defined
via a thermodynamic analysis, alternative definitions are used. The dynamics
of SOC systems occurring in chain reaction or ``avalanche'' like events, a set of
observables characterizing the avalanches,  size (s), duration (t), area (a), \etc 
 are defined.
 Fix such an  observable,
 say $N$,
 and compute the related 
probability $P_L(n) \deq  Prob(N=n)$
 at stationarity for a system of characteristic size $L$.
 The numerical simulations show that the graph of $P_L(n)$ exhibits  a power law 
behavior
over a finite range, with a cut-off corresponding to finite size effects. As $L$
increases the power law range increases. This  leads to the  conjecture that
as $L\to \infty$,
$P_L(n)$ converges
to a probability distribution $P^*(n)$, with a power law tail having
an exponent $\tau_n$
 called the {\it critical exponent} of the observable $n$. 
It seems commonly admitted in the SOC community
that a classification of the models can be made by the knowledge
of their  critical exponents (``universality classes'').
 Consequently, a large effort has been
devoted to the computation of these quantities.

Considerably less efforts have been made 
to establish a clear foundation of the basic SOC concepts and
to clarify  their connection to their classical statistical mechanics
counterpart \cite{BT,Sornette}. Clearly, this is a hard task since 
even preliminary notions like ``state'' or ''thermodynamic limit'',
though intuitively clear, suffer from a lack of  precise
mathematical definition in this setting. In discrete automata
like the BTW model \cite{Dhar} the state is the unique ergodic
probability measure, $\mu_L$, 
of a discrete Markov chain, finite when $L$ is finite.
 In continuous dynamical systems like the Zhang model \cite{Zhang}
there exists typically infinitely many ergodic measures and therefore
one has to add additional constraints to define the state $\mu_L$ 
in a non ambiguous way
\cite{BCK1,BCK2,BCK3}. The probability distributions
$P_L$, for the observables $a,s,t \dots$
 are directly obtained from $\mu_L$ \cite{BCK1,BCK3} but they contain 
less information. The observable $a,s,t$ are simply indicators of the
dynamics. There is no a priori reason to believe that
the knowledge of $P_L(s),P_L(a),P_L(t)$ or,
even, of the joint probability $P_L(a,s,t)$ gives all the relevant (that is, allowing
to classify the models into universality classes) informations about
the state $\mu_L$. 

The thermodynamic limit $L \to \infty$ and the supposed ``convergence''
of $\mu_L$ to a ``critical state'' poses  deeper problems
since even the proof that 
there exists \textit{indeed} a limiting state and that 
the probability
distribution of avalanche observables are still 
defined in this limit remains to be done, for most models. 
The usual classical
statistical mechanics constructions of the thermodynamic limit
like the Dobrushin-Landford-Ruelle's
(DLR)  
\cite{RuelleSM} cannot be directly applied because of the a priori absence of
a Gibbs formalism. 
 On the other hand, the methods used in interacting particle systems,
allowing to define the dynamics in the thermodynamic limit, 
 on the basis of the Hille-Yoshida theorem and the properties of Feller processes, 
requires locality conditions which are broken in SOC model. One has then to develop
new ideas for non-Fellerian Markov processes and this has only been done in a few
examples \cite{Maes}.

But one of the main problem is \textit{the treatment of the data obtained from 
finite size systems simulations  itself}
and the extrapolation to the $L \to \infty$ limit. Indeed, though it was believed in the
earlier SOC papers that this extrapolation  can be handled by classical 
finite-size scaling \cite{Kadanoff},
further investigations proposed alternative scaling \cite{Tebaldi,Vespignani1,Lise} and,
at the moment, there is no agreement on which scaling form applies. Consequently,  a lot of efforts have 
yet to be devoted to  the understanding and analysis of SOC models.\\

Though the analogy between self-organized criticality and usual critical phenomena is the core of the
SOC paradigm, it is remarkable 
that, up to now, some well developed techniques
of analysis of critical phenomena have not been adapted to
the study of SOC models. A phase transition has different manifestation.
It is in particular  characterized by a singularity 
of the thermodynamic
potential (free energy, pressure). At a phase transition point,
and for suitable interactions, the free energy, which is the generating function
of the cumulants, exists in the thermodynamic
limit but it is not analytic (in a $k$ th order phase transition
it is $\cC^{k-1}$ but not $\cC^k$). In many examples, 
the failure of analyticity is manifested
by the Lee-Yang phenomenon \cite{LeeYang}. For finite size systems the partition
function is a polynomial in a variable $z$ which typically depends
on control parameters like the temperature or the external field.
Since all coefficients are positive there is no zero on the positive real axis.
However, in the thermodynamic limit, at the critical point, some zeroes pinch the real
axis at $z=1$, leading to a singularity in the free energy.
The finite-size scaling properties of the leading zeroes and of the density
of zeroes near to $z=1$ determine the order of the transition \cite{Janke}
and also the critical exponents in the case of a second order phase transition
\cite{Kim}.

A natural question is
whether there exists a similar property in SOC,
namely can we exhibit a' ``free-energy like'' function, developing
singularities in a similar way in the infinite lattice size limit.
Though there exists a huge literature about the Lee-Yang zeroes,
there is, to the best of our knowledge, no attempt to study Self-Organized
Criticality from this point of view.
In this paper, we show that the cumulants generating function
of the probability distribution of the
observables $a,s,t, \dots$ have this property. More precisely,
the expected   convergence of $P_L$ to a power
law induces  a  Lee-Yang phenomenon for the corresponding
cumulants generating function (eq. \ref{ZL}).
We show that this effect is related to the observed divergence of
the moments. Furthermore, a scaling theory of the Lee-Yang
zeroes is proposed. 

After some preliminaries
(section \ref{GEN}), we give explicit analytical results (section \ref{THEO})
in several  cases
used as guidelines for subsequent analysis of a SOC model
(section \ref{ZHANG}). 
We first study  the truncated power law case where the cut-off tends to infinity 
when a parameter
$L$, (corresponding  the lattice size in SOC models),
tends to infinity  (section \ref{POW}). We give in particular an analytic expression
for the zeroes.
Then, we investigate the effect of a smooth cut off (section \ref{CO}).
We first discuss the properties that this cut off must have,
extrapolated from numerical simulations, and present some of the ansatz
found in the literature (section \ref{HypPl}). We then explicitly 
compute the Lee-Yang zeroes 
for a probability distribution obeying the 
finite-size scaling form proposed in \cite{Kadanoff}
and converging to a power law as $L \to \infty$ (section \ref{FSS}). 
We show in particular that,
when the power law exponent $\tau$ is larger that $1$ there is a \textit{violation}
of the scaling usually observed in classical critical phenomena; namely,
there is an anomalous logarithmic dependence on $L$ 
for the angle that the zeroes do with the real axis in the $t=\log(z)$
plane. We also show that when $\tau>1$ a bias
is artificially induced by the numerical simulations,
when the size of the sample used to generate the empirical probability
distribution is fixed independently of the lattice size.
This effect can be analyzed with the Lee-Yang zeroes (section \ref{cutoff}).
We then briefly
study some other scaling form proposed in the literature
and the effect of a finite size scaling violation
on the Lee-Yang zeroes
(section \ref{PAC}).
Finally, in the last section, we present
numerical simulations for the Lee-Yang zeroes in the Zhang SOC model
and compare them to the theoretical results obtained
in section \ref{THEO}. We see no clear cut evidence of finite size scaling
violation, but show that this model is quite sensitive to the
numerical cut off induced by a lattice size independent sampling. 
This can raise some doubts on the conclusions
about  scaling (Finite Size Scaling, multifractal
or whatsoever) which can be drawn from some large lattice simulations
done on this model in the literature.

\su{Probability distribution and Lee-Yang zeroes.}\label{GEN}

\ssu{The finite size system.}

Let $P_L(n)=Prob(N=n)$ be the probability distribution of the avalanche observable
$N \in 1 \dots \xi_L$, where the index $L$ refers
to the characteristic size of the system.  
 $\xi_L$, the maximal value that $N$ takes  is \textit{finite},
whenever $L < \infty$, but diverges
as $L \to \infty$.  Therefore, the  function:

\beq \label{ZL}
Z_L(z)=\sum_{n=1}^{\xi_L} z^nP_L(n)
\eeq

\nid where $z \in \bbbc$, is a polynomial of degree $\xi_L$.
In particular, since $Z_L(z)$ is an analytic function
of $z$ in the complex plane, all its moments exist. 
Denote
by $E[]_L$ the expectation with respect to  $P_L(n)$. Call :

\bea
m_L(q) = \sum_{n=1}^{\xi_L}P_L(n)n^q \deq E[n^q]_L
\eea

\nid where $q$ is a real (positive) number.
For integer $q$, the $m_L(q)$'s are the moments of $P_L(n)$. 
 Note that the normalization of $P_L(n)$ imposes $Z_L(1)=m_L(0)=1$.

For finite $L$, $Z_L(z)$ has $\xi_L$ zeroes in $\bbbc$,
that are either real $\leq 0$, or complex conjugate.
Denote them
by $z_L(k), k=1 \dots \xi_L$ and order them such that
$0 < |z_L(1)-1| \leq \dots \leq |z_L(k)-1| \leq \dots \leq |z_L(\xi_L)-1|$.
Note that $z=0$ is a trivial zero, of multiplicity one, since $P_L(1)>0$.
Write $z_L(k)=R_L(k)e^{i\theta_L(k)}
=1+r_L(k)e^{i\nu_L(k)}$.   
Since all $P_L(n)$  are positive, $Z_L(z)$ has no zero
on the positive real axis for finite $L$. Consequently, the log-generating
function\footnote{There is obviously a formal analogy
between (\ref{ZL}) (resp. (\ref{GL})) and a partition function (resp.
a free energy).} $log[Z_L(z)]$ is well defined on  $\bbbr^*_+$. Furthermore :

\beq \label{GL}
G_L(t) \deq log[Z_L(e^t)]
\eeq

\nid is an analytic function of $t$. Call :

\beq \label{chiLq} 
\chi_L(q)=\left. \frac{d^q}{d z^q} log[Z_L(z)]\right|_{z=1}
\eeq

\nid where $z=e^t$.
The quantities $\chi_L(q)$ 
 are easily expressed in terms of the Lee-Yang zeroes :

\beq \label{chiL}
\chi_L(q)
= (-1)^{q-1} (q-1)! \sum_{k=1}^{\xi_L}
\frac{1}{(1-z_L(k))^q}
\eeq

\ssu{The ``thermodynamic'' limit $L\to \infty$.}

\sssu{Divergence of the moments and Lee-Yang phenomenon.} \label{DIV}

As already written in the introduction,  a mathematical definition
of the thermodynamic limit in SOC is a difficult task, beyond the scope
of this paper. However, in \cite{BCK5},
 we developed a dynamical system approach for  the
Zhang model. Then, the thermodynamic formalism
\cite{Sinai,Ruelle,Bowen,Keller} can be used to define the finite size SOC state of
as a Gibbs measure\footnote{The particular structure of the Zhang model
allows to symbolically encode the dynamics. In the framework of the thermodynamic
formalism a Gibbs measure is a probability measure weighting the
symbolic chains encoding the trajectories
with an exponential weight called a potential (see \cite{BCK5} for details).}
 in this setting. 
It is then shown that the joint avalanche size distribution,
for example, can be obtained in this formalism via a proper
potential. The corresponding 
generating function for the time correlations, called the
topological pressure, is
the formal analog to the free energy.
In this setting,
it is argued that the  critical behavior expected
 in the thermodynamic limit, is manifested by 
a non analyticity of the topological pressure
as $L \to \infty$, that can be linked to the loss of hyperbolicity characterizing
the limit $L \to \infty$ of the Zhang model \cite{BCK4}.
The loss of analyticity can be easily detected by looking
at the generating function (\ref{ZL}). Indeed, its zeroes exhibit a Lee-Yang
phenomenon. 

The paper \cite{BCK5} is devoted to dynamical system aspects and to the mathematical
foundation of a thermodynamic formalism for the Zhang SOC model, and the link
between the scaling theory of Lee-Yang zeroes in classical critical phenomena
and general SOC model is not addressed.
This is the aim of the present paper. The results developed here
are therefore complementary to \cite{BCK5} but are independent.

The present paper focus on the analytic properties of the log generating 
functions of the probability of avalanches indicators, when  $L \to \infty$.
 It intends to analyze the variations in the Lee-Yang zeroes properties if
one uses the different scaling forms found in the SOC literature.
 Consequently, we collected
 the minimal implicit assumptions used in the SOC literature 
 and we infer the consequences
they lead to. This means that the result developed a priori hold for
all SOC models. \\
 
It is first
assumed that $P_L(n)$ converges to some  probability distribution
$P^*(n), \ n=1 \dots \infty$. It is furthermore assumed that
$P^*(n)$ has a power law tail\footnote{Note that the limiting probability distribution
is defined only if $\tau>1$.}, namely $P^*(n)=\frac{K}{n^\tau}$,
for a certain $n$ range, $n=n_0 \dots \infty$
where $n_0<\infty, \forall L$. The number $n_0$ depends on the model and on the
observable and introduces an extra parameter in the characteristics
of the probability distribution.
In the computations done in this paper there is no loss
of generality is assuming that $n_0=1$.  Therefore, in the sequel, $P^*(n)$
will stand for the power law $\frac{K}{n^\tau}, \ n=1 \dots \infty$.

The measured exponent $\tau$ in SOC  belongs to the interval $]1,2[$.
 $K=P^*(1)$ is the normalization constant. 
Consequently,  $K=\frac{1}{\zeta(\tau)}$
where $\zeta$ is the Riemann $\zeta$ function
\footnote{In general the normalization constant depends on $n_0$.}.
 Under the above assumptions,  the moments 
$m_L(q)$ behaves asymptotically like $\sum_{n=1}^{+\infty} n^{q-\tau}$.
This sum 
 diverges for all $q >\tau-1$. It is numerically observed that
$m_L(q)$ diverges like $m_L(q) \sim L^{\sigma(q)}$.
A central issue is to compute the \textit{scaling exponents }
given by: 

\beq
\sigma(q) \deq \lim_{L \to \infty} \frac{\log(m_L(q))}{\log(L)}
=\lim_{L \to \infty} \frac{\log(\chi_L(q))}{\log(L)}
\eeq

\nid  $\sigma(q)$ is an non decreasing function.
Its Legendre transform is  found under the name of  ``multifractal spectrum''
in the SOC literature \cite{Tebaldi}
though it has no direct connexion with the fractal geometry
of the invariant set.

 Since $P^*(n)$ is a probability
distribution the limiting generating function:

\beq
Z^*(z) = \lim_{L \to \infty}Z_L(z)=\sum_{n=1}^\infty P^*(n)z^n
\eeq

\nid is still an analytic function in the open unit
disc in $\bbbc$. However, 
the log-generating function of $P^*(n)$ is not analytic near to $z=1$
since the derivative of order $q>\tau-1>0$ diverge.
The corresponding singularity is related
 to the behavior of the zeroes in the vicinity of
$z=1$. More precisely,  fix $\epsilon >0$ arbitrary small,
 call $I_L(\epsilon)=\left\{i \ | \ |z_L(i)-1| < \epsilon \right\}$
and $n_L(\epsilon)=\#I_L(\epsilon)$ where $\#$ denotes the cardinality of a set.
Then the divergence of $\chi_L(q)$ is governed by the zeroes
which accumulate in $I_L(\epsilon)$. Namely, the sum (\ref{chiL}) contains
a singular term :

\beq \label{gammas}
 \gamma_s(L,\epsilon,q) =
(-1)^{q-1} 2.(q-1)! \sum_{k=1}^{\frac{n_L(\epsilon)}{2}}
 \frac{\cos(q\nu_L(k))}{r_L^q(k)}
\eeq

\nid which diverges as $L \to \infty$, while the remaining part in the sum is regular
and is bounded by $\frac{(q-1)!}{\epsilon^q}$ as $L \to \infty$.

\sssu{Scaling of the zeroes in classical critical phenomena.}

In the theory of classical critical phenomena,  it is
possible to relate the scaling exponents of quantities
such as magnetization or latent heat, susceptibility, 
etc ... to the behavior of the Lee-Yang zeroes near to $z=1$.
There exists a scaling theory based on earlier works
by Lee and Yang \cite{LeeYang}, Grossmann and Rosenhauer 
\cite{Grossmann}, Abe \cite{Abe}, Suzuki \cite{Suzuki},
Privman and Fisher \cite{Fisher},
Itzykson et al. \cite{Zuber}, Glasser et al. \cite{Privman}. 
Many analytical and  rigorous results are
also known (for example \cite{Newman,Sokal,Biskup,RuelleGL}). A lot
of efforts have been devoted to the study of ferromagnetic systems
(e.g. Ising or Potts models) though many other examples have also been
studied in the literature. 
In this setting, one distinguishes
the zeroes in the complex magnetic field (called Lee-Yang zeroes)
from the zeroes in the temperature plane (Fisher zeroes). In the
first case, the zeroes lie on the unit circle for a large
class of models including the Ising's one.

The Fisher zeroes usually approach
$t=0$ in the $t=\log(z)$-complex plane with a constant angle $\phi$
(this is the case for the Ising model and mean-field
ferromagnetic models \cite{Privman}) .
 This allows to obtain simple scaling
expression for the singular part of the free energy $f^s(t)$
where $t$ is the reduced temperature. In this setting,  
an analytic expression for $f^s(t)$
has been obtained by Grossmann and Rosenhauer 
\cite{Grossmann}, and, later on, extended by
Itzykson et al. \cite{Zuber} by using the renormalization
group theory. This approach  has been extended by Glasser et al. \cite{Privman} to mean-field
models. In the thermodynamic
limit $f_s^{\pm} \sim A_\pm|t|^{2-\alpha}$
where $A_\pm$ are universal constants ($\pm$ label the
two magnetic phases at low temperature), and $\alpha$
is the critical exponent for the specific heat.
It follows from the renormalization group
analysis \cite{Fisher} that the singular part 
of the free energy obeys a  Finite-Size Scaling form:

\beq \label{scalingfreeenergyclas}
f^s(t,V)= \frac{1}{V} \cF[t(AV)^{\frac{1}{2-\alpha}}]
\eeq

\nid where  $V$ is the finite volume
and $\cF$  a universal function. Accordingly, the $n$ first
 Fisher zeroes are given by :

\beq \label{normalscal}
t_V(n)=\left[
\frac{2\pi}{\left(A_+^2+A_-^2-A_+A_-\cos(\pi\alpha) \right)}  
\frac{n}{V}\right]^{\frac{1}{2-\alpha}}
e^{i(\pi-\phi)} 
\eeq

 The angle $\phi$ is related to $A_\pm,\alpha$
by $\tan\left[(2-\alpha)\phi\right]=\frac{\left[\cos(\pi\alpha)-\frac{A_-}{A_+}\right]}{\sin(\pi\alpha)}$.
This situation, where the zeroes  approach
the singularity
 with a \textit{constant angle} $\phi$ and where the modulus scales like
the volume to a certain power will be referred to as \textit{normal scaling} in the sequel.

It seems a general observation \cite{Grossmann} that the zeroes lie on a
curve or a union of curves dividing the complex plane in different
regions of analyticity of $Z^*(z)$, corresponding to different phases.
More precisely, it has been recently proved by Biskup et al. \cite{Biskup}
 that the zeroes lie
on curves with a simple analytic expression and accumulate in the thermodynamic
limit on loci where the various branches of the free energy have the
same modulus. This last result suggests that a wide extension of the Lee-Yang
phenomena can be made toward  dynamical systems near to a critical point.
 We now develop this aspect for the analysis of the log generating function
of probability distribution in the SOC framework.  In the sequel, we will
not distinguish between the Lee-Yang zeroes and the Fisher zeroes
and we will use the generic terminology ``Lee-Yang'' for the zeroes.  

\su{Scaling theory of SOC and Lee-Yang zeroes.} \label{THEO}

In this section, we establish analytical results for various
finite size scaling forms found in the SOC literature. These
results are then used in section \ref{ZHANG} for the analysis
of the empirical data obtained from  a numerical simulation of a SOC model.
As a matter of fact, for finite size SOC systems, the power law is
truncated by a cut off characterized by a length scale $\Lambda_L$,
usually different from $\xi_L$. The starting point is therefore
the analysis of a truncated power law with a sharp
cut-off at a value $\Lambda_L$. This
 is a useful example for subsequent analysis since
the analytic form of $P_L(n)$ and the cut-off is known.

\ssu{Zeroes of a truncated power law.} \label{POW}

Assume that $P_L(n)=\frac{C_L}{n^\tau},
n=1 \dots \xi_L$ where $C_L$ is a normalization constant
and $\xi_L=\Lambda_L=L^\beta, \tau >1,\beta>0$. Furthermore, assume
that $\tau < 2$.

\sssu{Scaling of the moments and log generating function.}

For $0\leq q \leq\tau-1$,
$m_L(q) \to \frac{\zeta(\tau-q)}{\zeta(\tau)}$ and consequently $\sigma(q)=0$.
The non-zero scaling exponents $\sigma(q)$ can be obtained 
from the following integral approximation of $m_L(q)$, which becomes
exact in the limit $L \to \infty$, provided $q \geq \tau-1$ : 

\beq \label{approxmLq}
m_L(q) \sim C_L\Lambda_L^{q+1-\tau} 
\int_{\frac{1}{\Lambda_L}}^{1}
 u^{q-\tau}du = \frac{C_L}{q+1-\tau}\left(\Lambda_L^{q+1-\tau} -1 \right)
\eeq

Then,  $\sigma(q)=\beta(q+1-\tau)$ for (real) $q>\tau-1$.
Note however, that for finite size, one has additional $L$ dependent
terms which have to be considered when extrapolating
from numerical simulations.
It is also interesting to note that  formula (\ref{approxmLq}) 
gives 
useful informations on the \textit{rate of convergence} of $m_L(q)$ to a constant
for $q<\tau-1$. Indeed, the convergence 
 is \textit{not uniform} in $q$, namely
the \textit{closer} is $q$ to $\tau-1$ the \textit{slower} is the convergence rate.
This means that a \textit{systematic bias due to finite size}
is introduced in the numerical
simulations when dealing with the  $q$'s close to $\tau-1$. This produces
a spurious curvature, near to $\tau-1$,  for the function $\sigma(q)$ extrapolated from
numerical data. This effect, which disappears as $L\to \infty$, can lead
to misleading conclusion since it can be
interpreted as an evidence of a multifractal scaling. \\ 

The scaling exponents can also be obtained from the scaling
of the log generating function. Indeed, the generating function writes:

\beq \label{ZLpuis}
Z_L(t) = 1 + \sum_{n=1}^{\xi_L} P_L(n)(e^{tn}-1)   
\sim 1 + C_L\Lambda_L^{1-\tau} \int_{\frac{1}{\Lambda_L}}^1 
 u^{-\tau}(e^{t\Lambda_Lu}-1)du 
\eeq

Set $t'=\Lambda_L t$ and 

\beq \label{psipow}
\psi(t') \deq \int_{0}^1 u^{-\tau}(e^{t'u}-1)du
= \sum_{n=1}^\infty \frac{t'^n}{n!}\int_{0}^1 u^{n-\tau}du
=\sum_{n=1}^\infty \frac{t'^n}{n!(n+1-\tau)}
\eeq

Note that since $\tau < 2$, this integral is finite as
can easily be checked by integration by part.
Therefore the commutation of the integral and the series is allowed.

Consequently,  

\beq \label{fpartitionpow}
Z_L(t) \sim 1 + C_L \Lambda_L^{1-\tau}\psi(t') = 1+ C_L L^{\beta(1-\tau)}\psi(tL^\beta)
\eeq

\nid and :

\beq 
G_L(t) = \log(1+C_L L^{\beta(1-\tau)}\psi(tL^\beta))
\eeq

$\psi(t)$ is a smooth function of $t$ which vanishes as $t \to 0$.
Therefore, for $t \to 0$ :

\beq \label{freeenergypow}
G_L(t) \sim C_L L^{\beta(1-\tau)}\psi(tL^\beta)
\eeq

\nid which gives the right scaling for the moments by differentiating with respect
to $t$ at $t=0$. One remarks that this scaling form
is analogous to the form (\ref{scalingfreeenergyclas}).

\sssu{Lee-Yang zeroes.}

From equation (\ref{fpartitionpow}) the zeroes are approximately given by:

\beq \label{fpartitionzeroespow}
\psi(t')=-C_L \Lambda_L^{\tau-1}=-C_L L^{\beta(\tau-1)}
\eeq

Since  $\tau > 1$, $\Lambda_L^{\tau-1}$
diverges. $\psi(t')$ is an increasing
function of the real variable $t'$
which vanishes as $t'=0$ and tends to $-\infty$
when $t' \to -\infty$.
Furthermore, for any $K >0$,
$\psi(t')$ is bounded by $\psi(K)$
in the ball  $|t'|<K$ in the complex plane.
Consequently, the equation (\ref{fpartitionzeroespow}) can
be fulfilled only if $t'_L(k)$'s have
\textit{a diverging modulus as $L$ grows}. On  the other
hand, since $t_L(k) = \frac{t'_L(k)}{\Lambda_L}$
converges to zero, $|t'_L(k)|$ must  grow slower
than $\Lambda_L$. It grows in fact like
$\log(\Lambda_L)$ as shown below.

 Note that, conversely, when $\tau < 1$, 
$-C_L L^{\beta(\tau-1)}$ goes to zero in the thermodynamic limit
\footnote{Though the limit probability is not defined it is
nevertheless possible to investigate  the finite size scaling properties for the zeroes of
the finite size generating function.}.
Then, the zeroes are formally given by :

$$t_L(k)=\frac{1}{\Lambda_L}\psi_k^{-1}(-C_L \Lambda_L^{\tau-1})
\sim\frac{1}{\Lambda_L}\psi_k^{-1}(0)$$

\nid where $\psi^{-1}_k$ is the $k$ th branch of the inverse of $\psi$
in the complex plane. Consequently, the zeroes have a simple scaling in these
case similar to (\ref{normalscal}).
 In particular, \textit{the argument does not depend on $L$}.\\

The observed values of the critical exponents in SOC, $\tau > 1$,
induces anomalous scaling that can be observed on the Lee-Yang zeroes.
Though the form (\ref{fpartitionpow}) can be used to compute the Lee-Yang zeroes,
 it is easier to use: 

\beq \label{ZLpow}
Z_L(t) = \sum_{n=1}^{\Lambda_L} P_L(n)e^{tn}   
\sim  C_L\Lambda_L^{1-\tau} \int_{\frac{1}{\Lambda_L}}^1 
 h(u,t')du 
\eeq

\nid where  $h(u,t')=u^{-\tau}e^{t'u}$.
There exists several techniques to compute the Lee Yang zeroes in statistical
mechanics.
A standard way is to argue that the asymptotic free energy admits
different analytic continuation in different  regions of the complex 
plane, separated by Stokes lines where the zero accumulate
in the thermodynamic limit. Indeed, because of the large number of terms in the
polynomial which make up the partition function, the behavior tends to become
dominated by some set of the coefficients. Thus we have different analytic functions
in different regions of the complex plane. These functions have oscillating phases
but smoothly varying amplitude. The zeroes 
locates then on Stokes boundaries
where two types  of behavior have comparable magnitude \cite{Zuber,Privman,Biskup}.
The Stokes boundaries becomes cuts in the thermodynamic limit.
Across the boundaries the free energy has a regular real part and jumps
in the imaginary part  \cite{Grossmann}.

Applying this strategy to our formal partition function
(\ref{ZLpow}), one identifies easily two regions.
For real $t'$, as $u$ grows from to $0$ to $\infty$,
$h(u,t')$ first decay like $u^{-\tau}$ until
a minimum $u_-= \frac{\tau}{t'}$. Therefore,
$u_- > \frac{1}{\Lambda_L}$ when $t < \tau$.
For $u >u_-$,  $h(u,t')$ grows exponentially
like $e^{t'u}$. Therefore, when $t'$ is small
 the integral
in (\ref{ZLpow}) is essentially dominated by the algebraic decay $u^{-\tau}$
and $\int_{\frac{1}{\Lambda_L}}^1 h(u,t')du \sim 
 \frac{1}{\tau-1}\left[\Lambda_L^{\tau-1} -
1\right]$.
On the other hand, for large $t'$, $u_- \to 0$ and the algebraic part is negligible
compared to the exponential part. Hence  $\int_{\frac{1}{\Lambda_L}}^1 h(u,t')du 
\sim \int_{u_-}^{1} e^{t'u}du = \frac{1}{t'}\left[e^{t'}-e^\tau\right]$. 
This argumentation extends to complex
$t'$ and
suggests that one can roughly divide the $t$ complex
plane into two regions where $Z_L(t)$ has a different analytic form:
 for sufficiently small $t'$ the algebraic
part dominates, while for large $t'$ the exponential part is dominant. 
Then the zeroes have to stay at the place where the two forms are
of the same order. Therefore an approximate equation for
the location of the zeroes is given by:

\beq 
\label{Zeroscalpow}
\frac{1}{t'} \left[e^{t'} + a_1t' - a_2\right] 
 = -\frac{\Lambda_L^{\tau-1}}{\tau-1}$$
\eeq 

\nid where $a_1=\frac{1}{1-\tau}, a_2 =   e^\tau$.\\

The solutions of $\frac{e^{t'}}{t'}
 =  -\frac{\Lambda_L^{\tau-1}}{\tau-1}$ are 
given by :

\beq \label{SolPow}
t'_L(k)=\Lambda_L t_L(k)=-W_k(\frac{\tau-1}{\Lambda_L^{\tau-1}}) 
\eeq

\nid where $W_k(x)$ is the $k$ th branch of the Lambert function \cite{Knuth}.
Note that the Lambert function has infinitely many branch and consequently the
equation $\frac{e^{t'}}{t'}
 =  -\frac{\Lambda_L^{\tau-1}}{\tau-1}$ has infinitely many solutions.
Indeed, in replacing the initial sum by an integral in (\ref{ZLpow}) we have introduced
spurious zeroes which have to be removed for finite $L$. In the sum (\ref{ZLpow}) one has
a step of integration $\frac{1}{\Lambda_L}$ which defines  a lower cut-off
in the scales one has to consider.  Consequently,
only the branches $k= -\frac{\Lambda}{2} \dots \frac{\Lambda}{2}$, where one takes
into account the symmetry of the zeroes with respect to the real axis, are relevant.

The Lambert function $W_k$ admits the following expansion, for large 
$|\log(x)|$ \cite{Knuth}:

\beq \label{Lambert}
W_k(x)= \log_k(x) - \log(\log_k(x)) + \sum_{l\geq 0}\sum_{m\geq 1}c_{lm} 
\frac{\left(\log \log_k(x) \right)^m}{\left(\log_kx\right)^{l+m}}
\eeq

\nid where $\log_k$ is the $k$ th branch of the complex logarithm and
$c_{lm} =\frac{1}{m!}(-1)^l\left[^{l+m}_{l+1} \right]$ is expressed in
terms of the Stirling numbers of the first kind $(-1)^{m+n}\left[^{n}_{m}\right]$ 
\cite{Knuth}.

The double series $\sum_{l\geq 0}\sum_{m\geq 1}c_{lm} 
\frac{\left(\log \log_k(x) \right)^m}{\left(\log_kx\right)^{l+m}}$
is absolutely convergent for sufficiently large $|\log(x)|$ \cite{Knuth}.
 Since we are only interested in the asymptotic divergence when $L$
grows, 
one can therefore neglect the series in the asymptotic. Note however
that   the convergence to the asymptotic regime where the series becomes
negligible \textit{is faster when the product $\beta(\tau-1)$ is larger}.

The term $\log(\log_k(\frac{\Lambda^{\tau-1}}{\tau-1}))$ cannot be neglected
compared to $\log_k(\frac{\Lambda_L^{\tau-1}}{\tau-1})$ since it contains
crucial $k$ dependence for the real part of $t_L(k)$ (see below).
It is interesting to note that it introduces a $\log(\log L))$ finite size
scaling correction. A similar correction as been found in \cite{Sokal2}
for the Potts model with $q \geq 4$.

The corrections due to the other terms of eq. (\ref{Zeroscalpow})
become rapidly negligible as can easily be seen by a perturbation
expansion. \\

One finally obtains the following asymptotic form for the Lee
Yang zeroes of the truncated power-law:

\bea \label{zeroespow}
\Re(t_L(k)) &\sim& \frac{\left[-\log(\frac{\tau-1}{\Lambda_L^{\tau-1}}) +
\frac{1}{2}\log\left( \log^2(\frac{\tau-1}{\Lambda_L^{\tau-1}})
+4k^2\pi^2\right)
 \right]}{\Lambda_L} \label{tzeroespowR}\\
\Im(t_L(k)) &\sim& \frac{2k\pi}{\Lambda_L} -\frac{1}{\Lambda_L}
 \arctan\left(\frac{2\pi k}
{\log(\frac{\tau-1}{\Lambda_L^{\tau-1}})}\right)\label{tzeroespowangle}
\eea

The term 
$\log(\log_k(\frac{\tau-1}{\Lambda_L^{\tau-1}}))$  
introduces a $k$ dependence which
implies in particular that the zeroes (in the $z$ plane) do not lie
on circle but on a more complicated curve (see Fig. \ref{patternzeroespow}).
This dependence remains important, even for the first 
zeroes, 
up, to very large $L$, especially if $\tau$ is close to $1$.
Indeed, for a fixed $k$ the term $\log^2(\frac{\tau-1}{\Lambda_L^{\tau-1}})$
dominates the term $4k^2\pi^2$ only for $L >> \left((\tau-1)e^{2\pi k}
 \right)^\frac{1}{\beta(\tau-1)}$ (say, for $\tau=1.25,\beta=2.67$ this corresponds
to a $L >>1500$).  

The $\arctan$ term in the imaginary part 
acts essentially as a phase
term $ -\frac{1}{\Lambda_L} \arctan\left(\frac{2\pi k}
{\log(\frac{\tau-1}{\Lambda_L^{\tau-1}})}\right)$
 which is slowly varying  (in the $k$ variable) 
compared to the dominant term $\frac{2k\pi}{\Lambda_L}$.
Furthermore, since the $\arctan$ is bounded above
by $\frac{\pi}{2}$ it is rapidly negligible as $k$ grows. Therefore, one can 
consider with a good approximation that  
$\Im(t_L(k)) \sim \frac{2k\pi}{\Lambda_L}$.

The argument of  $t_L(k)$  formally corresponds to the angle  
that the Fisher zeroes do with the real axis in critical phenomena.
 For a power law with $\tau <1$
this angle is independent of $L$ as discussed above. Conversely, for $\tau >1$
it is given by :

\beq \label{Argpow}
Arg(t_L(k))\sim
\arctan\left(
\frac
{2k\pi}
{\left[-\log(\frac{\tau-1}{\Lambda_L^{\tau-1}}) +
\frac{1}{2}\log\left( \log^2(\frac{\tau-1}{\Lambda_L^{\tau-1}})
+4k^2\pi^2\right)
 \right]}
\right)
\eeq

One observes therefore a \textit{logarithmic deviation to the normal scaling on the 
argument of the $t_L(k)$'s}. The conclusion is therefore that, though the truncated
power law
obeys the classical \textit{finite size scaling} form (\ref{scalingfreeenergyclas}),
 the Lee-Yang zeroes
display nevertheless an \textit{anomalous scaling} due to the exponent
$\tau >1$ (see Fig. \ref{argtpow}).

In the $z$ plane the zeroes $z_L(k)=e^{t_L(k)}$
are approximately given by:

\bea \label{zzeroespowR}
R_L(k)&=&|z_L(k)| \sim 1+\Re(t_L(k))\\
\theta_L(k)&=&\Im(t_L(k)) \sim \frac{2k\pi}{\Lambda_L}
= \frac{2k\pi}{L^\beta}
\label{zzeroespowtheta}
\eea

Therefore the arguments $\theta_L(k)$ of the zeroes in the $z$ complex plane
 are  uniformly distributed in $[-\pi,\pi]$ with a good approximation. 

Finally, one can 
 determine
 the exponents $\tau,\beta$ from the Lee-Yang zeroes. 
The exponent $\beta$ corresponds to the scaling exponent 
of the correlation length $\xi_L$.
 The equation (\ref{zzeroespowtheta}) 
provides \textit{a straightforward way} to compute it. 
Furthermore, the exponent $\tau$
can be obtained from eq. (\ref{tzeroespowR}). The term $4\pi^2k^2$ 
certainly rapidly dominates the term $\log^2(\frac{\tau-1}{\Lambda_L^{\tau-1}})$
in  the modulus $R_L(k)$ as $k$ grows. This is a fortiori true for
$k \sim \Lambda_L$ which correspond to the zeroes the farthest from $z=1$.
Consequently:

\beq \label{RLtau}
R_L(\theta) \sim \left[\frac{\Lambda_L^{\tau-1}}{\tau-1}
\sqrt{\log^2(\frac{\tau-1}{\Lambda_L^{\tau-1}})
+4\theta^2\Lambda_L^2} \right]^\frac{1}{\Lambda_L} 
\sim \left[\frac{2\theta\Lambda_L^\tau}{\tau-1}\right]^\frac{1}{\Lambda_L}
\eeq

If one takes $k=\frac{\Lambda_L}{2}$ (corresponding
to an angle $\pi$), one has:

\beq
\tau=\lim_{L \to \infty}\frac{\Lambda_L \log(R_L(\pi))}{\log(\Lambda_L)}
\eeq 

\nid which allows a possible determination of $\tau$
from the scaling of the Lee-Yang zeroes. Note however that the convergence
is logarithmic in $L$.

\sssu{Numerical checks.}

Since it is easy to generate numerically a power law distribution one is a priori
free to choose any values for $\tau$ and $\beta$. 
However, the  closer  $\tau$ is to $1$ the slower
is the convergence to the asymptotic regime where the formulas obtained in the previous
section hold. More precisely, 
the rate of convergence is essentially governed by the product
$\beta(\tau-1)$. The closer is $\tau$ to $1$ the larger
has to be $\beta$. But the larger  $\beta$, the faster the degree of
the polynomial increases with $L$ and therefore the time needed
for the computation of the zeroes increases.
On the other hand, since the theory developed
here is independent of the actual value of $\beta,\tau$ (provided $\tau>1$),
 we mainly studied  examples where $\beta=2$ and $\beta=2.2$ which gives a reasonable increase
in the polynom degree, and $\tau=1.9$ such that the product $\beta(\tau-1) \sim 2$.

We first depicted the pattern of zeroes, for different sizes, in
 Fig. \ref{patternzeroespow}. 

\bef  
\bc  
\begin{minipage}{6cm}   
\epsfxsize=6cm  
\epsfysize=6cm  
\epsffile{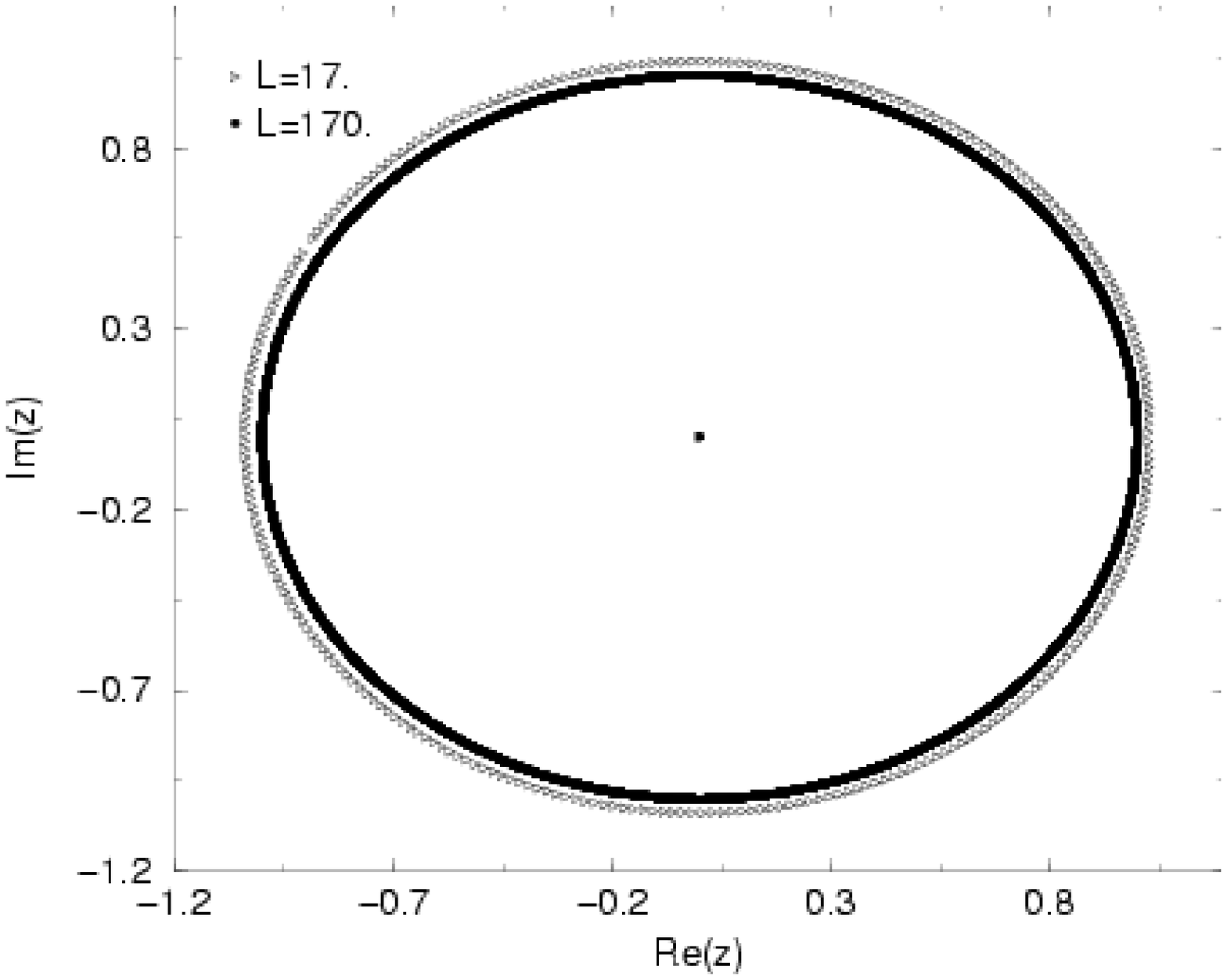} 
\end{minipage} \hspace{1cm} 
\begin{minipage}{7cm}   
\epsfxsize=6cm  
\epsfysize=6cm  
\epsffile{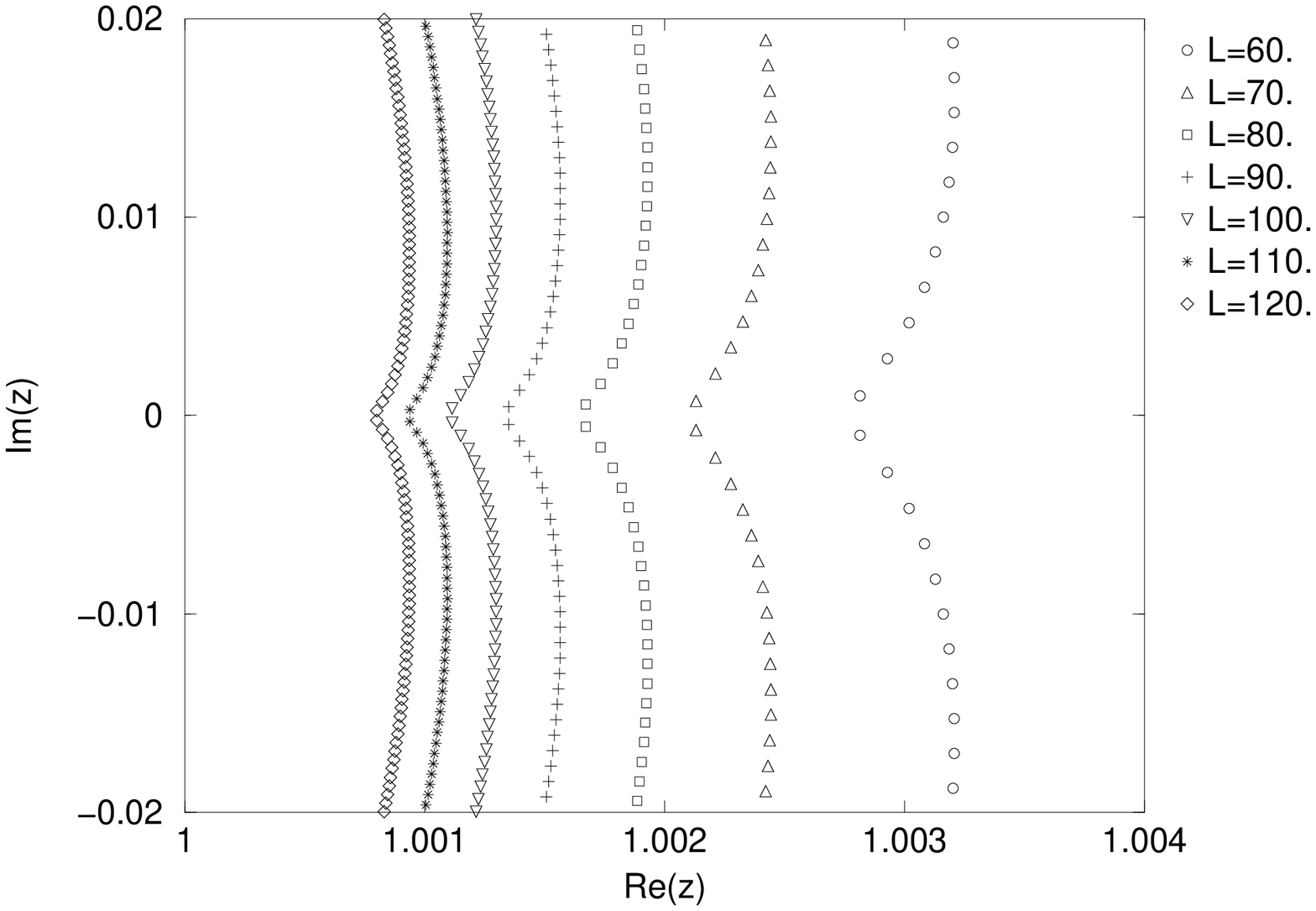} 
\end{minipage} \hspace{1cm} 
\vspace{0.3cm} 
\caption{ a:
Lee-Yang zeroes for various $L$ values
in the $z$ complex plane $\beta=2,\tau=1.9$.
Fig. \ref{patternzeroespow} b: Local behavior near to $z=1$.}
\label{patternzeroespow}
\ec
\enf   

One notices the slow convergence to the unit circle, and the shape of the curve
near to $z=1$: this is not a circle.

We plotted Fig. \ref{zeroestpow}
the real and imaginary
part of the zeroes in the $t$ plane.
 For the real part we tried
 a fit of the form $r(k)=\frac{\left(-\log(a)
+\frac{1}{2} \log\left[\log^2(a)+4\pi^2k^2 \right] \right)}{\Lambda_L}$
where $a$ is a free parameter. The formula (\ref{tzeroespowR}) gives $a_{th}
=\frac{\tau-1}{\Lambda_L^{\tau-1}}$
but, in the several approximations we made, we neglected some constants,
 and one expects
$a$ to be different from the theoretical value, with an error that should decrease
as $L$ grows. The result of the fitting is represented Fig. \ref{zeroestpow}a
 for the $200$ first
zeroes. We found indeed that the experimental value $a_{exp}$
 is closer and closer to its theoretical value
as $L$ increases and that our approximation is better and better as $L$ increases.
 For $L=50$,
$a_{th}= 7.87 \times 10^{-4}, a_{exp} = 5.6 \times 10^{-4} \pm 4 \times10^{-5}$;
for $L=100$, $a_{th}= 2.26 \times 10^{-4}, a_{exp} = 1.5 \times 10^{-4} \pm 3
 \times 10^{-5}$
and for $L=150$, $a_{th}= 1.09 \times 10^{-4}, a_{exp} = 7.1 \times 10^{-5} 
\pm 8 10^{-7}$.
The imaginary part is plotted fig. \ref{zeroestpow} 
b together with the theoretical prediction
$\Im(t_L(k))=\frac{2\pi k}{L^\beta}$ (straight lines). We remarked a slight deviation
of the first zeroes to $\frac{2 \pi}{\Lambda_L}$ due to the
correction term in (\ref{zzeroespowtheta}).

\bef  
\bc  
\begin{minipage}{6cm}   
\epsfxsize=6cm  
\epsfysize=6cm  
\epsffile{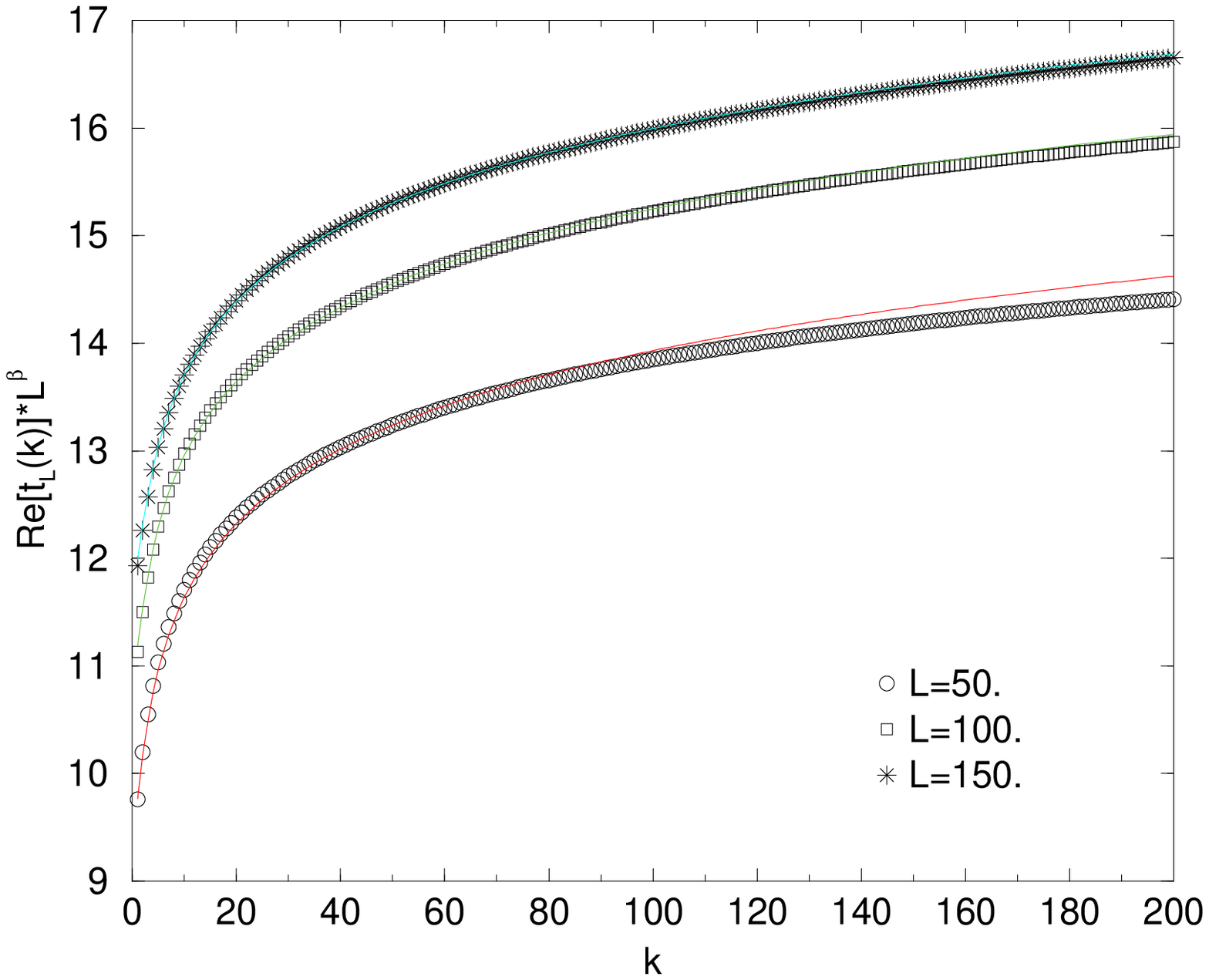} 
\end{minipage} \hspace{1cm} 
\begin{minipage}{6cm}   
\epsfxsize=6cm  
\epsfysize=6cm  
\epsffile{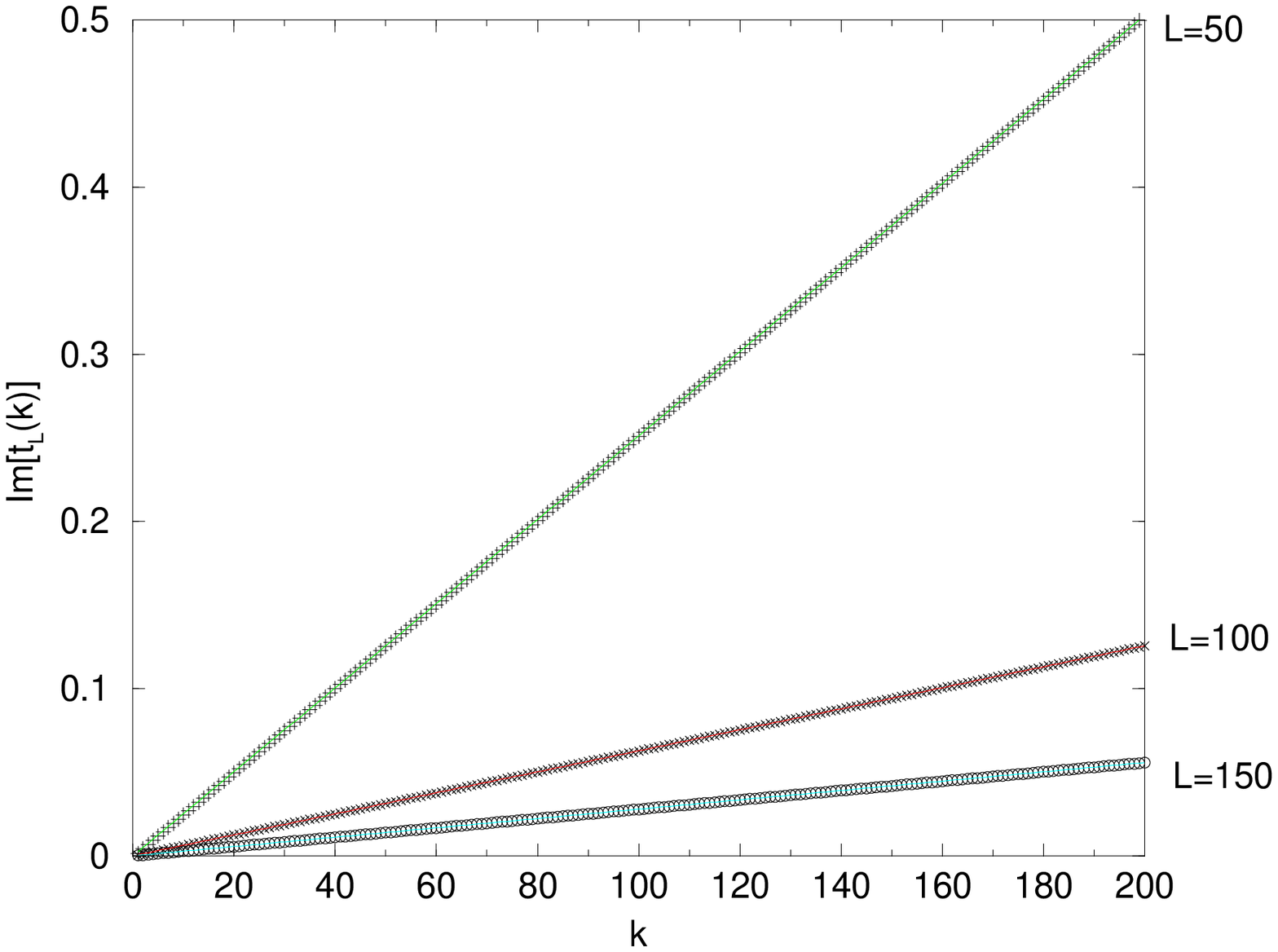} 
\end{minipage} \hspace{1cm} 
\vspace{0.3cm} 
\caption{ Lee-Yang zeroes in the $t$ complex plane, 
versus $k$ for various $L$ values. Fig. \ref{zeroestpow}a: Real part  multiplied
by $\Lambda_L$.
Fig. \ref{zeroestpow} b: Imaginary part. The fitting curves are plotted in color, 
full lines.}
\label{zeroestpow}
\ec
\enf   

Fig.  \ref{argtpow} a, we plotted the argument of $t_L(k)$ as a function of $L$ for $k=1 \dots 5$.
We notice the logarithmic deviation to the normal scaling  as
predicted by formula (\ref{Argpow}). We tried to fit these curves with a fit of the
form $\arctan\left(\frac{2k\pi}{\alpha(log(x)+\gamma)+ \frac{1}{2}
\log(\alpha^2(log(x)+\gamma)^2 + 4k^2\pi^2)} \right)$ where $\alpha,\gamma$
are free parameters. In fig.  \ref{argtpow} b, we show the different scaling occurring
for $\tau < 1$ (normal scaling) or $\tau >1$ (anomalous scaling).

\bef  
\bc  
\begin{minipage}{6cm}   
\epsfxsize=6cm  
\epsfysize=6cm  
\epsffile{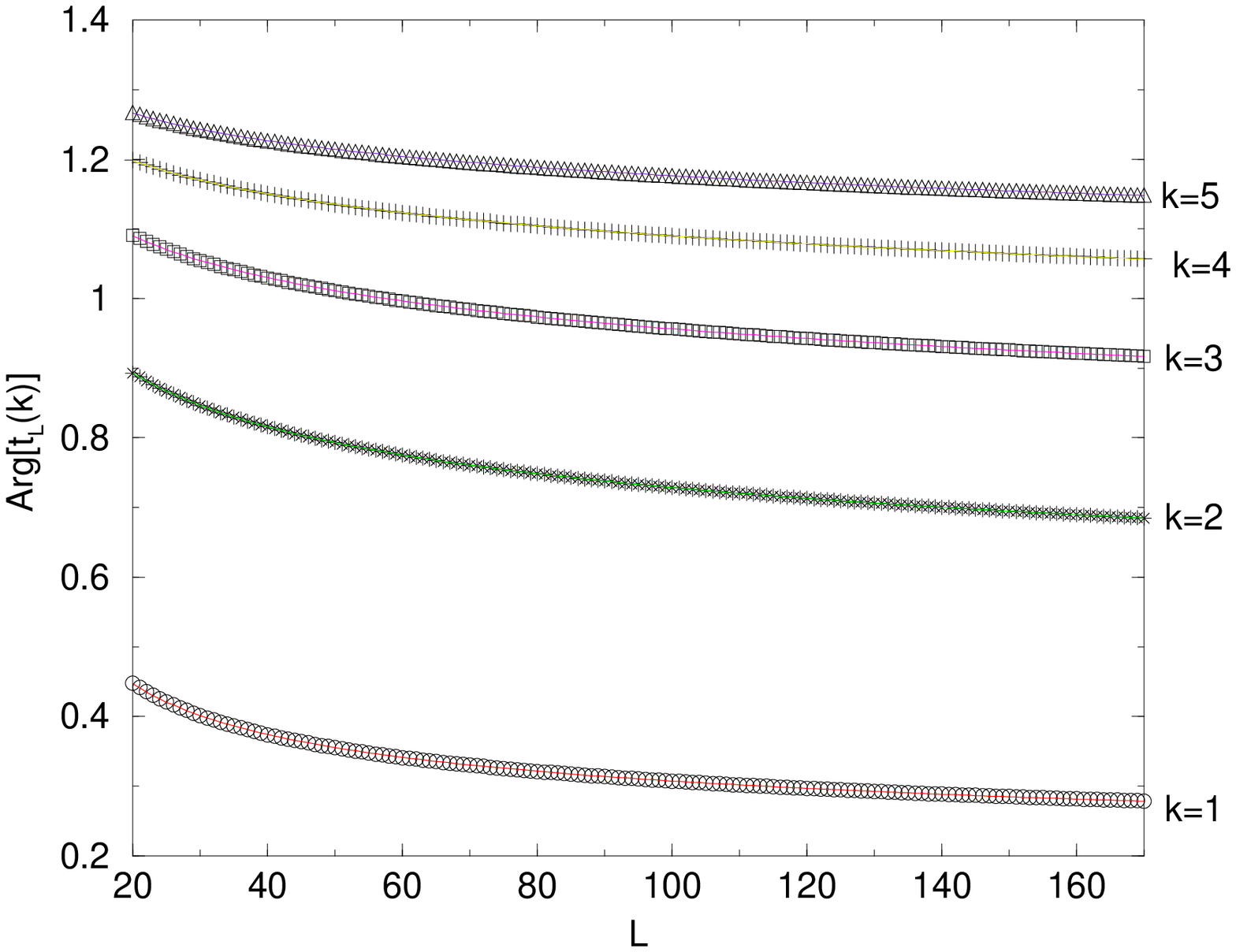} 
\end{minipage} \hspace{1cm}
\begin{minipage}{6cm}   
\epsfxsize=6cm  
\epsfysize=6cm  
\epsffile{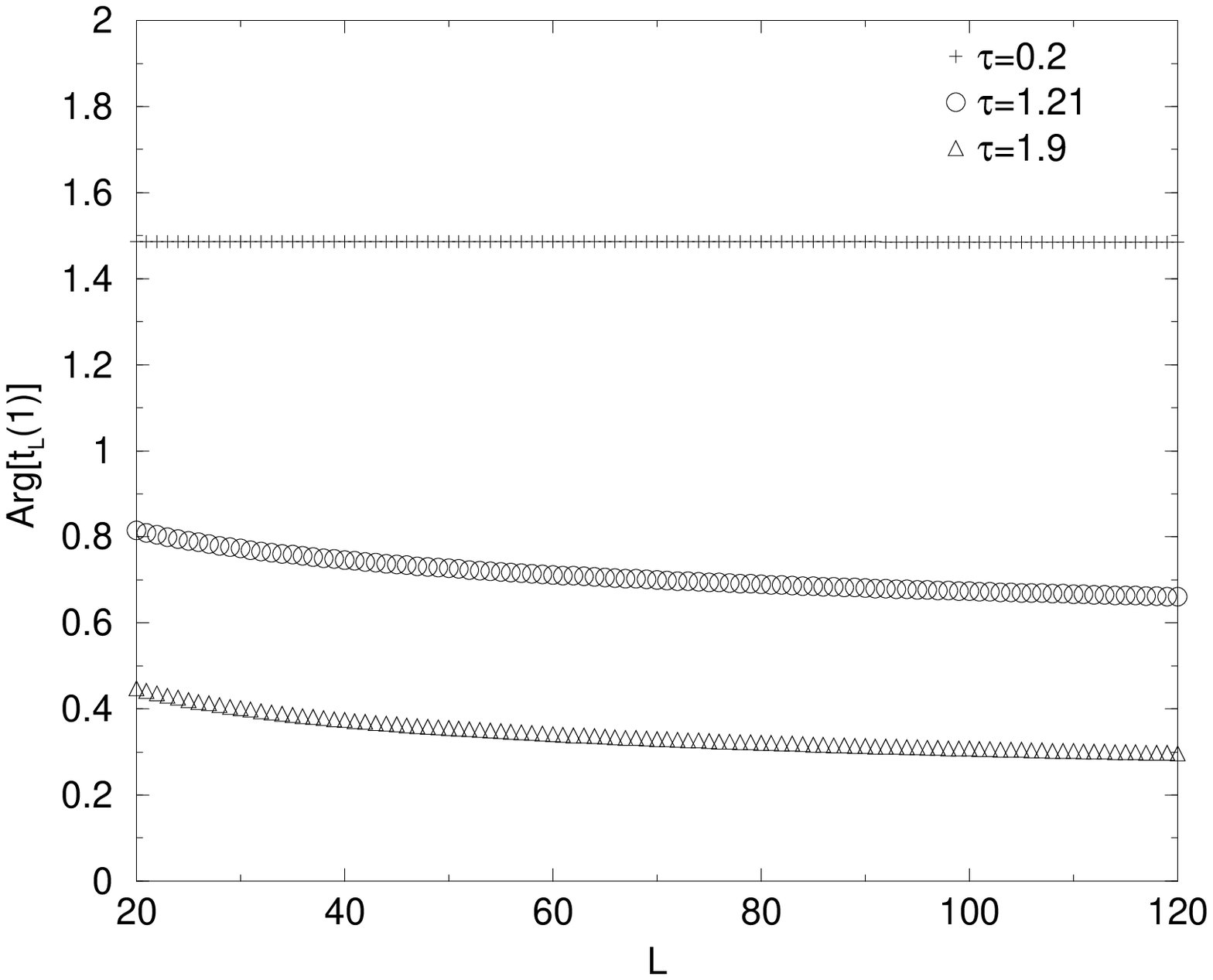} 
\end{minipage}
\vspace{0.3cm} 
\caption{a Argument of the 
Lee-Yang zeroes in the $t$ complex plane, 
versus $L$ for various $k$ values, $\beta=2, \tau=1.9$.
Fig. \ref{argtpow}b. Normal ($\tau=0.2$) and anomalous ($\tau=1.21,\tau=1.9$)
scaling of the angle with the real
axis for various $\tau$ value. The fitting curves are plotted in color, 
full lines}
\label{argtpow}
\ec
\enf   

Finally we argued above that the exponents $\beta$ and $\tau$ can be determined
with a good accuracy from the scaling of the zeroes. In fig. a we plotted
$Arg(z_L(5))=\theta_L(5)$ as function of $L$ for $\beta=1.5,\tau=1.6;\beta=2,\tau=1.21;
\beta=2.2,\tau=1.9$ and $L=20 \dots 120$.
We choose the $5$th zero rather than the first ones since for the first zeroes
the correction coming from the $\arctan$ term in eq. (\ref{tzeroespowangle})
 influences slightly
the scaling for small $L$. We tried a fit of the form $\frac{C}{x^\beta}$
where $C_{th} \sim 10\pi=31.415\dots$.
 The fits shown on
the figure gave respectively : $\beta=1.507 \pm 0.002; \beta=2.007 \pm 9 \times
10^{-3};
\beta=2.205 \pm \times 0.001$ 

For the determination of $\tau$ we used eq. (\ref{RLtau}) for $\tau=0.2;1.21;1.9$
and $\beta=2$. We have interpolated the data with a fit form 
$e^{\beta\tau\log(ax)/x^\beta}$
where $a,\tau$ were free parameters. For $L=20 \dots 120$,
we found respectively  $0.1998(2) \pm 6.10^{-6}$,$1.207(5) \pm 7.10^{-5}$,
$1.89(5) \pm 0.0001$ for the value of $\tau$. This
 is  satisfactory especially when taking into
account the smallness of the $L$'s we considered. 

\bef  
\bc  
\begin{minipage}{6cm}   
\epsfxsize=6cm  
\epsfysize=6cm  
\epsffile{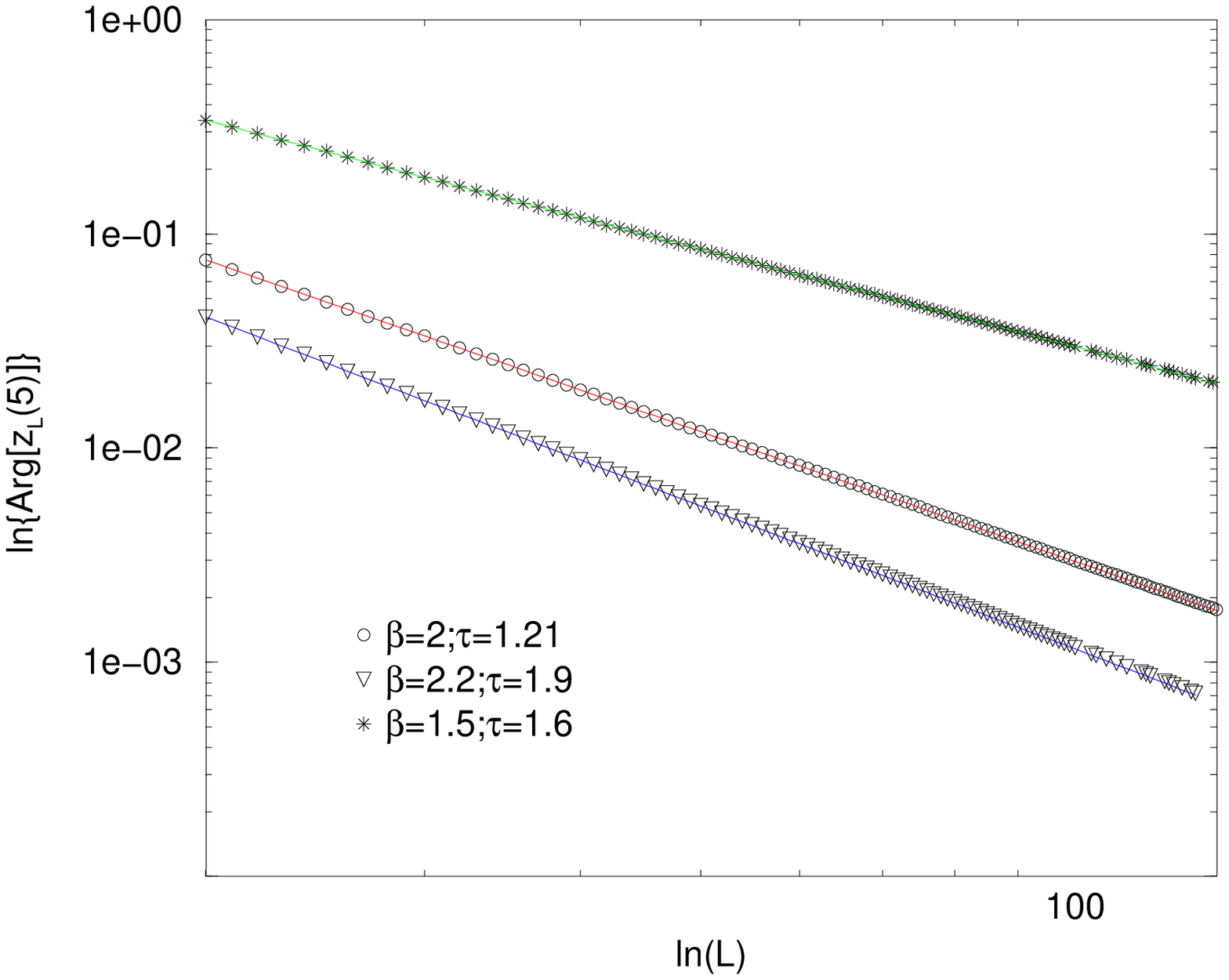} 
\end{minipage} \hspace{1cm}
\begin{minipage}{6cm}   
\epsfxsize=6cm  
\epsfysize=6cm  
\epsffile{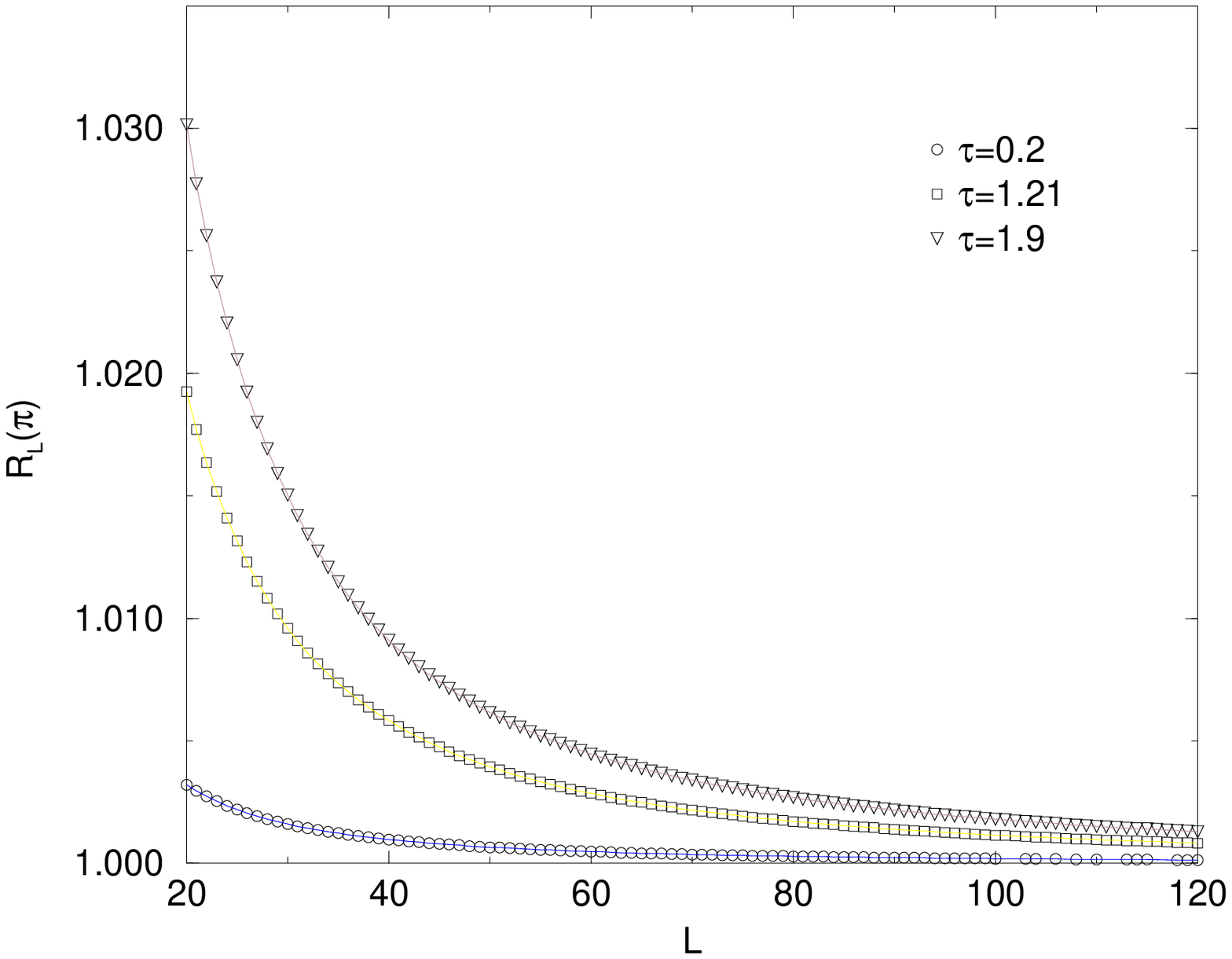} 
\end{minipage}
\vspace{0.3cm} 
\caption{a Argument of the 
fifth Lee-Yang zero in the $z$ complex plane, 
versus $L$ for various $\beta$ values.
Fig. \ref{argzpow}b. $R_L(\pi)$ as a function of $L$ for various $\tau$
values. The fitting curves are plotted in  color full lines.}
\label{argzpow}
\ec
\enf

\ssu{General case : the structure of the cut off.} \label{CO}
 
The observed probability distribution of avalanches observables
in SOC models is in general a power-law truncated by a cut-off function associated
to finite-size effects. Except for a few case \cite{RuelleSen}, the
analytic form of the cut-off is not known. Consequently, various
scaling form have been proposed in the SOC literature. In this section
we grasp the most common scaling forms and discuss their effect on the Lee-Yang
zeroes. We also investigate the effect of the sampling cut-off inherent
to numerical simulations.
 
\sssu{General assumptions about the cut off function.} \label{HypPl}

 Since $P_L(n)$ is expected to converge to
a power law $K n^{-\tau}$ one can write, without loss of generality,
the finite size probability under the form:

\beq \label{PLfL}
P_L(n)=\frac{C_L}{n^\tau}f_L(n), \ 1 \leq n\leq \xi_L
\eeq

\nid where   $C_L$ is a normalization constant depending
on $L$.
  $f_L(n)$ is the finite size cut-off.

The graph obtained from numerical simulations
suggests that $f_L(n)$  is a regular function which obeys:

\bea
\nonumber (i)   && \  \lim_{L \to \infty} f_L(n) =1, \ \forall n \leq \xi_L. \\
\nonumber (ii) && \  \forall p > 0, \lim_{n \to \infty} n^p f_L(n) = 0, \ \forall L  
\eea

\nid The property (i) corresponds to the pointwise convergence
to a power law.
(ii)  characterizes the observed
fact that the tail
of $P_L(n)$ decreases faster than any power of $n$
(e.g. is  least exponentially decreasing).

These properties are not sufficient for a scaling
theory and further assumptions have to be made.
We now discuss  various scaling form that one can find
in the literature but also the numerically induced cut-off
effects.\\

\textit{Finite-Size Scaling.}
In 1990, Kadanoff et al. \cite{Kadanoff} proposed a Finite-Size Scaling Ansatz 
where :

\beq \label{Kada}
f_L(n)=g(\frac{n}{\Lambda_L})
\eeq

\nid  $g$ being a universal function.
$\Lambda_L=L^\beta$ is the characteristic scale for a lattice of size $L$ and
 $\xi_L=\alpha\Lambda_L$ where $\alpha$ is a constant.
The case of a truncated power law developed in section \ref{POW}
corresponds to the particular case where $\alpha=1$, and $g(x)$
is equal to $1$ for $x \in [0,1]$ and zero otherwise.
Note that in this case the property (ii) has the consequence that:
$$ (iii) \ \lim_{x\to 0} g(x) = 1$$

\textit{multifractal scaling.} In the same paper, Kadanoff et al. \cite{Kadanoff}
discussed another form of scaling, that is:

\beq 
\frac{\log(P_L(n))}{\log(\frac{L}{L_0})} = 
f\left(\frac{\log(\frac{n}{n_0})}{\log(\frac{L}{L_0})}\right)
\eeq

\nid where $f$ is universal and does not depend explicitly on $L$.
$L_0,n_0$ are some constants that can be omitted by a suitable
redefinition of the quantities, then:

\beq \label{KadaMF}
P_L(n)=L^{f(\log n/log L)}
\eeq

 This  representation is called by the authors
an $f-\alpha$ representation, $\alpha$ being the quantity 
$\frac{\log(\frac{n}{n_0})}{\log(\frac{L}{L_0})}$.  In this case one has a whole
spectrum of scaling indexes, i.e. all the values taken on
by $\frac{df}{d\alpha}$. In the general case $f$ is non linear.
Then the universality class is given by
the function $f$, rather than by a finite set of critical exponents.
In this
case, the scaling exponents are non linear functions of $q$. \\

\textit{Finite Size Scaling violation and convergence to
a power law.} 
Another scaling form  has been introduced by Lise and Paczuski \cite{Lise}
from numerical simulations on the OFC model \footnote{B.C. is very grateful
to M. Paczuski for illuminating discussions on this topic in Bielefeld.}.
Lise and Paczuski analyzed their data with the following form for $P_L(n)$:

\beq \label{PPac}
P_L(n) =C_L n^{-\tau}L^{F_L\left(\frac{\log(n)}{\log(L)}\right)}; \ n=1 
\dots L^{\beta_L}
\eeq

\nid where $\beta_L$ is now $L$ dependent,
 $\beta_L < \beta < \infty$
and $\beta_L \to \beta$ as $L \to \infty$.
Furthermore,
the numerical plot of $F_L(x)$ in \cite{Lise} suggests that 
$F_L(x)$ converges to a ``step'' function $Y(x-\beta)$ as $L \to \infty$, 
where $Y(u) = 0, \ u \in ]-\infty,0[$,
$Y(0)=C$
and $Y(u)=-\infty, \  u \in ]0,\infty]$.
The finite size scaling case corresponds to $F_L(x)=
\frac{\log(g(L^{x-\beta}))}{\log L}$ where $g$ is defined in (\ref{Kada}).
 This example is quite interesting since
it gives an example of a probability distribution violating
(\ref{Kada}) but converging nevertheless to a power law (namely the
exponents $\tau$ and $\beta$ are still meaningful). We remark indeed
that the corresponding probability distribution is
\textit{not multifractal}
in the sense of \cite{Tebaldi} since it is easy to check 
that the scaling exponents are given by
$\sigma(q)=\beta(q+1-\tau),\forall q\geq\tau-1$ and $\sigma(q)=0,
\forall q<\tau-1$.
This case is therefore intermediate between the Finite-Size Scaling 
(\ref{Kada}) and the multifractal case (\ref{KadaMF}).\\

In the following sections we discuss the behavior of the Lee-Yang zeroes
in these different cases.

\sssu{Finite-Size Scaling.} \label{FSS}

 We show in this
section that the behavior of the Lee-Yang zeroes in the finite
size scaling case is essentially the same as for the truncated power law, 
provided $g$ in (\ref{Kada}) fulfills the conditions (i),(ii),(iii)
in the previous section.

It is well known that the moments obey the same scaling.
This can be recovered from the generating function (\ref{ZL}). 
 Provided $ \tau < 2$ it
 scales
as $L \to \infty$ like :

\beq
Z_L(t) \sim 1 + C_L\Lambda_L^{1-\tau} \left[\Upsilon_\alpha(t') -
\Upsilon_{\frac{1}{\Lambda_L}}(t')\right] \sim 
1 + C_L\Lambda_L^{1-\tau}\Upsilon_\alpha(t') + C_L \psi(t) \sim
1 + C_L\Lambda_L^{1-\tau}\Upsilon_\alpha(t')
\eeq

\nid  for $t \to 0$ in the $t$ complex plane. In this equation $t'=\Lambda_L t$ and :

\beq
\Upsilon_\gamma(t') \deq \int_{0}^\gamma u^{-\tau}g(u)(e^{t'u}-1)du
= \sum_{n=1}^\infty \frac{t'^n}{n!}\int_{0}^\gamma u^{n-\tau}g(u)du
\eeq

Consequently, near to $t=0$,
$Z_L(t) \sim 1 + C_L L^{\beta(1-\tau)}\Upsilon_\alpha(tL^\beta)$ and 
$G_L(t) \sim C_L L^{\beta(1-\tau)}\Upsilon_\alpha(tL^\beta)$. As expected one
obtains the same scaling as for a truncated power law. The only change is
that  the  function $\Upsilon$ depends now on the $g$ function.\\

 The zeroes of $Z_L(t)$ are therefore well approximated by :

\beq\label{Zerokada}
\Upsilon_\alpha(t') \sim -C_L\Lambda_L^{\tau-1}
\eeq

\nid which is completely analogous to (\ref{fpartitionzeroespow}).
 Therefore, the same conclusions hold:
the equation (\ref{Zerokada}) can
be fulfilled only if the solutions, $t'_L(k)$, have
a diverging modulus as $L$ grows. The computation of the
zeroes is essentially the same as in section (\ref{POW})
which slight complications due to the presence of the
function $g$.

One can write :

\beq\label{ZLFSS}
Z_L(t)   
\sim  C_L\Lambda_L^{1-\tau} \int_{\frac{1}{\Lambda_L}}^\alpha
 h(u,t')du 
\eeq

\nid where now $h(u,t')=u^{-\tau}g(u)e^{t'u}$.
Recall that, by hypothesis, $g(u)$ is an at least exponentially
 decreasing function.
 As $u$ grows from to $0$ to $\infty$,
$h(u,t')$ first decay like $u^{-\tau}$ until
a minimum $u_-$ after which $h(u,t')$ grows exponentially
like $e^{t'u}$. $u_-$ tends to $0$ as $t' \to \infty$.
More precisely, if $t'$ is sufficiently large,  from (iii) $g(u)$ is essentially $1$
on the interval $[0,u_-]$ and therefore
$u_-$ is approximately given by $t'u_-^{-\tau} = \tau u_-^{-\tau-1}$.
 Consequently $u_- \sim \frac{\tau}{t'}$.
 When $u$ is sufficiently
large the decay coming from $g(u)$ compensates the exponential
increase of $e^{t'u}$. Therefore, there is a maximum $u_+$
after which $h(u,t')$ tends to zero with a rate given
by $g(u)$. Here, $h(u,t')$ is essentially
$g(u)e^{t'u}$.  Therefore $u_+$ is given by
$t'g(u_+) + g'(u_+)=0$ or $t'=-\frac{d\log(g(u_+))}{du}$,
where $-\frac{d\log(g(u))}{du}$
is a function which increases faster than $u$
by $(ii)$.
Consequently, $u_+$ diverges as $t' \to \infty$.
Since $\alpha$ is bounded by assumption,
 $u_+ > \alpha$ for $t'$ (resp. $L$) sufficiently large.
But, from eq. (\ref{Zerokada}) we know that the modulus of the $t'$
corresponding to the zeroes
diverge as $L \to \infty$.
Consequently, provided $L$ for is sufficiently large, the zeroes
lie in a region where $u_+ > \alpha$. Guided by the wisdom coming
from section \ref{POW} we can assume that the zeroes accumulate onto a curve
$\gamma_L$ which separates the complex plane into two regions.
In the first one, the behavior is dominated by the algebraic part and
the integral in (\ref{ZLFSS}) is essentially
$\int_{\frac{1}{\Lambda_L}}^{\alpha}h(u,t')du 
\sim \frac{1}{\tau-1}\left[\Lambda_L^{\tau-1}
 -\alpha^{\tau-1}\right]$.
In the second region, the exponential part dominates.
Provided $t'$ is sufficiently large ($u_+ >> \alpha$),
the variations of $u^{-\tau}g(u)$ are small compared
to the increase of $e^{t'u}$ and this function can be
approximated by some  constant $\Gamma_g$. 
Hence  $\int_{u_-}^{\alpha} h(u,t')du \sim \Gamma_g\int_{u_-}^{\alpha}e^{t'u}du
\sim \frac{\Gamma_g}{t'} \left[e^{t'\alpha}  - e^\tau \right]$.

Consequently, for sufficiently large $L$, the zeroes are well approximated by

\beq\label{Zeroscal}
\frac{1}{t'} \left[e^{t'\alpha} + a_1t' - a_2\right]
 = -\frac{\Lambda_L^{\tau-1}}{\Gamma_g(\tau-1)}$$
\eeq

\nid where $a_1= \frac{\alpha^{\tau-1}}{\Gamma_g(\tau-1)}, a_2 = e^\tau$.
  
This equation is  similar to eq. (\ref{Zeroscalpow}). The zeroes are now given by :

\beq \label{SolKada}
t_L(k)=- \frac{W_k({\frac{\alpha\Gamma_g(\tau-1)}{{\Lambda_L}^{\tau-1}}})}{\alpha\Lambda_L}
\eeq

Consequently, one finds that the pattern of zeroes in the finite-size scaling case
is essentially the same as the power law case, up to a correction
depending on $\alpha, \Gamma_g$. More precisely, using the series expansion
(\ref{Lambert})
of the Lambert function one finds :

\bea \label{tzeroesFSS}
\Re(t_L(k)) &\sim& \frac{\left[-\log(\frac{\alpha\Gamma_g(\tau-1)}
{\Lambda_L^{\tau-1}}) +
\frac{1}{2}\log\left( \log^2(\frac{\alpha\Gamma_g(\tau-1)}{\Lambda_L^{\tau-1}})
+4k^2\pi^2\right)
 \right]}{\alpha\Lambda_L}\\
\Im(t_L(k)) &\sim& \frac{2k\pi}{\alpha\Lambda_L}
\eea

\nid where $k=1 \dots \xi_L=\alpha\Lambda_L$.

In the $z$ plane this essentially results in  a trivial re scaling of the argument $\Im(t_L(k))$
and a slight change in the modulus
$R_L(k)$ :

\beq \label{scalingFSSPow}
R_L^{FSS}(k) \sim \kappa_L
R_L^{PL}(k)^\frac{1}{\alpha}
\eeq

\nid where the superscript FSS (resp. PL) refers to the Finite Size scaling (resp.
Power Law) situation.
$\kappa_L=\left(\frac{1}{\alpha\Gamma_g} \right)^\frac{1}{\alpha \Lambda_L}$
is therefore a scaling factor which tends to $1$ as expected since the zeroes
have to accumulate on the unit circle. In equation (\ref{scalingFSSPow}) 
we neglected the more complicated dependence coming from the $\log(\log)$ term.
This has a small effect on the first zeroes but becomes negligible as $k$ grows.
This provides a way to determine $\kappa_L$. Setting $R_L^{PL}(\pi)$ for the
farthest zero from $z=1$:

\beq \label{kappaL}
\kappa_L = \frac{R_L^{FSS}(\pi)}{R_L^{PL}(\pi)^\frac{1}{\alpha}}
\eeq 

The argument of $t_L(k)$'s is  :

\beq \label{argtFSS}
Arg(t_L(k))\sim
\arctan\left(
\frac
{2k\pi}
{\left[-\log(\frac{\alpha\Gamma_g(\tau-1)}{\Lambda_L^{\tau-1}}) +
\frac{1}{2}\log\left( \log^2(\frac{\alpha\Gamma_g(\tau-1)}
{\Lambda_L^{\tau-1}})
+4k^2\pi^2\right)
 \right]}
\right)
\eeq

The cut off $g$
modifies therefore the value of the angles but \textit{not the scaling}.\\

We numerically checked these result in the following case. We generated
a probability distribution given by : 

\beq \label{ExFSS}
P_L(n)=C_L n^{-\tau}g(\frac{n}{L^\beta}) ; \ n = 1 \dots \alpha L^\beta
\eeq

\nid where :

\beq \label{gExFSS}
g(x)=e^{-x^\gamma}
\eeq

 We fixed the parameters to the values:
$\beta=2,\tau=1.9,\gamma=2.5,\alpha=1$. We show Fig. \ref{mappattern}
the collapse of
the curve of zeroes to the corresponding curve  for
a  power law with the same $\tau,\beta$. $\kappa$ was computed
from the ratio (\ref{kappaL}). We found $\kappa=1.0001$ for $L=100$.

\bef
\bc  
\begin{minipage}{6cm}
\epsfxsize=6cm  
\epsfysize=6cm
\epsffile{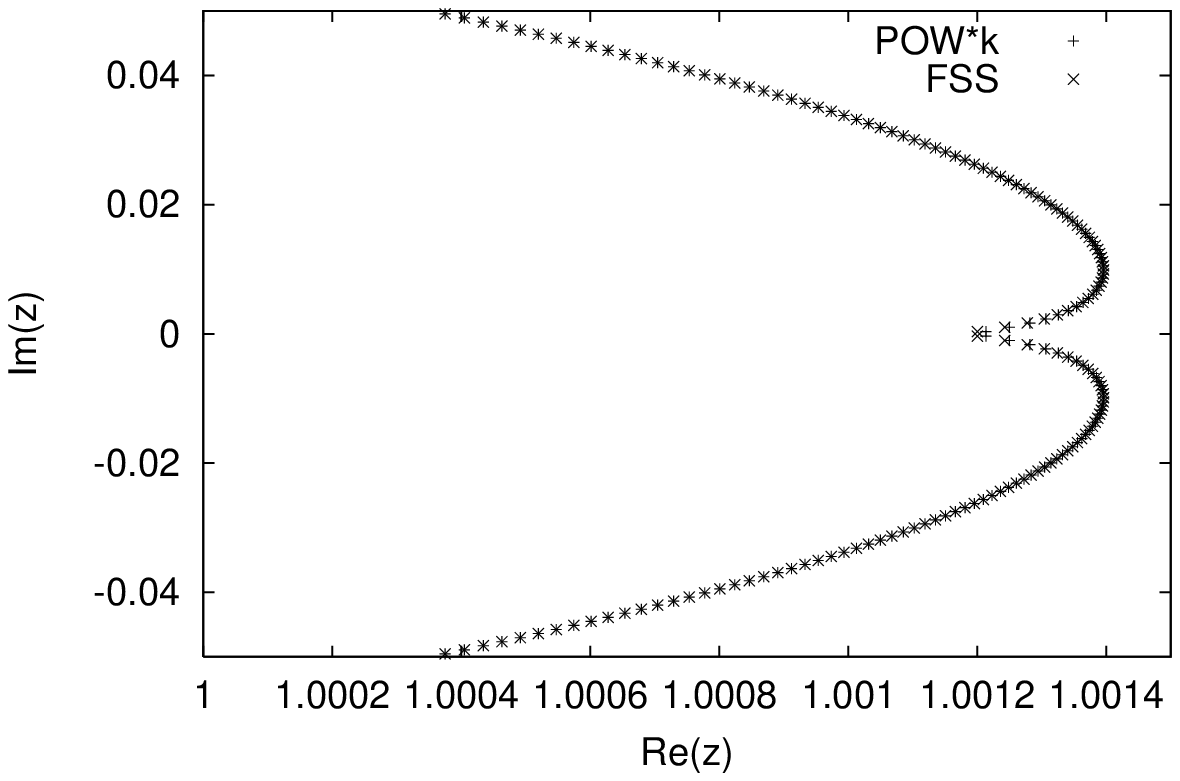} 
\end{minipage}
\vspace{0.3cm} 
\caption{ Pattern of zeroes in the $z$ complex plane
for the power law case, $\beta=2,\tau=1.9,L=100$,
 where the real and imaginary part have been
multiplied by $\kappa$, and for the FSS case 
$\beta=2,\tau=1.9,\gamma=2.5,\alpha=1,L=100$.}
\label{mappattern}
\ec
\enf

\sssu{Effect of a sampling cut off.} \label{cutoff}

We would like now to point out
 a very simple way to  violate the  scaling form
(\ref{Kada}),
by the only numerical procedure traditionally
used in the computation of $P_L(n)$. This effect is closely related
to the anomalous scaling of the Lee-Yang zeroes since it appears
for a critical exponent $\tau >1$.

In numerical simulations,
one computes an empirical distribution
$P^{exp}_L(n,\omega) = \frac{\cN(n,\omega)}{\cN}$
where $\cN$ is the total number of avalanche observed during the simulation
and $\cN(n,\omega)$ is the number of times an avalanche of size $n$ was observed.
This number is a random variable depending, say, on the initial condition(s),
or more generally,
on the seed $\omega$ used in the random generator.
However, one expects the system to be ergodic in a strong sense
such that  $P^{exp}_L(n,\omega) \to P_L(n)$ as $\cN \to \infty$ for
generic choices of $\omega$
(see \cite{BCK3} for details.).
However, since $\cN$ is finite, there exists wild fluctuations
 in the tail of the distribution. Furthermore,
 the avalanches such that $P_L(n) < \frac{1}{\cN}$
have a small (though non zero \footnote{However, since the largest non zero value of
$P^{exp}_L(n,\omega)$
is $\frac{1}{\cN}$,  the events such
that $P_L(n) << \frac{1}{\cN}$, even when observed, are given an incorrect
probability by the numerical procedure. The discrepancy with the theoretical
value increases as $n$ increases.\label{fn}})  probability to be observed
in a numerical simulation and this probability decreases as $n$ increases.
Obviously, there are several methods
such as smoothing or binning, allowing to reduce the
effects of noise in the tail. 

There exists however another, more subtle effect.
In all the examples of numerical computations
that we have found in the SOC literature, the value
of $\cN$ is fixed, \textit{independently of the system size}. This induces 
a pathological bias affecting the extrapolation to the thermodynamic limit
whatever the method used to analyze the empirical distribution.
In particular a violation of finite-size scaling can be observed
on $P^{exp}_L(n,\omega)$ \textit{even if the theoretical probability
$P_L(n)$ obeys (\ref{Kada})}. When $\cN$ is kept fixed
while $L$ increases, the estimation of 
the maximal value $\xi_L$ that
the random variable can take (defining the exponent $\beta$) 
is more and more biased. Indeed, while the true $\xi_L$ diverges as $L \to \infty$,
the empirical value $\xi_L^{exp}(\omega)$ converges to a constant.
Consequently, the probability distribution extrapolated to the thermodynamic limit
from the empirical distribution is biased.  
The aim of this section is to
analyze this effect, not discussed in the literature. \\

Assume therefore that $P_L(n)$ is
like (\ref{PLfL}) with a cut off obeying (\ref{Kada}).
Call $F_L(n)=\sum_{k=1}^n P_L(k)$ the corresponding repartition
function. Assume now that we perform
a finite sampling of the probability distribution with $\cN$ 
trials $X_1, \dots X_\cN$ where $X_i, i= 1 \dots \cN$
are independent \footnote{We assume here
that the trials are independent for simplicity. In SOC models, the avalanches
are not independent though, for finite $L$, the correlation decay can be
fast (it is exponential in the finite size Zhang model).}, 
identically distributed random variable,
with probability $P_L(n)$. Assume furthermore that $\cN$ is fixed independently
of $L$.
 Call 
$\xi_{L,\cN}^{exp}=\max\left\{X_k, \ k=1 \dots \cN\right\}$
the maximal value observed in the finite sampling.
The repartition function of the random variable $\xi_{L,\cN}^{exp}$ is
$F_L^\cN(x)$. Its average is given by:

\beq \label{Moymax}
E[\xi_{L,\cN}^{exp}]=\xi_L-\sum_{n=1}^{\xi_L}F_L^\cN(n)
\eeq
 
Clearly, were $L$ fixed while $\cN \to \infty$, then would
$E[\xi_{L,\cN}^{exp}]$ converges to $\xi_L$.
In fact the ergodic theorem gives a stronger statement, namely
$\xi_{L,\cN}^{exp} \to \xi_L$ almost-surely in this case. This essentially means that
for sufficiently small $L$, $E[\xi_{L,\cN}^{exp}]$ gives
a good estimate of $\xi_L$. 
On the other hand, if $\cN$ is fixed while $L$ increases, the correction term
$\sum_{n=1}^{\xi_L}F_L^\cN(n)$ in (\ref{Moymax}) becomes more and more
important leading to a wrong estimation of $\xi_L$. To be more precise
fix a value $y \in ]0,1[$,  such that $F_L^\cN$ is considered
as non negligible as soon as $F_L^\cN(n)>y$. Hence
$y$ is somehow arbitrary here (say close to $0$).
 Since $F_L^\cN(n)$ is strictly increasing  the equation $F_L^\cN(x)=y \Leftrightarrow 
F_L(x)=y^\frac{1}{\cN}$ has a unique solution $x_L \equiv x_L(y)$,
$\forall y \ \in \ ]0,1[$.
 If $L$ is small
(or if $\cN$ is sufficiently large) $x_L \sim \xi_L$ and therefore
$F_L^\cN(n)$ is essentially non zero for $n \sim \xi_L$. In this case
the term $\sum_{n=1}^{\xi_L}F_L^\cN(n)$ in (\ref{Moymax})
is negligible compared to $\xi_L$. 

On the other hand, $x_L$ is bounded from above by a \textit{finite} value $x$ such
that $F^*(x)=K\sum_{n=1}^{x} n^{-\tau} = y^\frac{1}{\cN}$, where
$F^*$ is the repartition function of the limiting probability $P^*(n)$.
 Consequently, as $L$ increases, for any $y \in ]0,1[$, $\frac{x_L(y)}{\xi_L}
\to 0$. Hence, the
function $F_L^\cN(x)$ is non negligible  on a larger and larger 
interval, whose length scales like $\xi_L$. Therefore,
the sum  $\sum_{n=1}^{\xi_L}F_L^\cN(n)$ in (\ref{Moymax}) becomes more
and more important, yielding a decrease in the expectation of the empirical maximum.

In the range of $L$ values where this effect starts to manifest
one has $x_L \sim \xi_L$. Hence, 
$\sum_{n=1}^{\xi_L}F_L^\cN(n) \sim \sum_{n=x_L}^{\xi_L}F_L^\cN(n)$
and $F_L^\cN(n)$ has only to be estimated in the interval $[x_L,\xi_L]$.
Furthermore, $F_L(n)=1-\sum_{k=n}^{\xi_L} P_L(n)
\sim 1-C_L\int_{n}^{\xi_L}
u^{-\tau}g(\frac{u}{\Lambda_L})du$. When $x_L$ is close to  $\xi_L$,
$\int_{n}^{\xi_L}
u^{-\tau}g(\frac{u}{\Lambda_L})du \sim (\xi_L -n) \xi_L^{-\tau}g(\alpha)$
for $n \in [x_L,\xi_L]$.
Furthermore,
in this range, $1-C_L\xi_L^{-\tau}(\xi_L -n) g(\alpha)$
is small compared to $1$. Hence, $F_L^\cN(n)
\sim 1-\cN C_L\xi_L^{-\tau}(\xi_L -n) g(\alpha)$.
The equation $F_L^\cN(x)=y$ has therefore
an approximate solution $x_L =\xi_L -
\frac{1-y}{\cN C_L g(\alpha)}\xi_L^{\tau}$.
 Then $\sum_{n=1}^{\xi_L}F_L^\cN(n)
\sim \sum_{n=x_L}^{\xi_L}F_L^\cN(n)$
can be roughly approximated by a linear interpolation giving
$\sum_{n=x_L}^{\xi_L}F_L^\cN(n)\sim \frac{\xi_L-x_L}{2} =\frac{1-y}{\cN C_L g(\alpha)}\xi_L^{\tau}$.

Consequently, the empirical expectation is, in this approximation:

\beq \label{approxxi}
E[\xi_{L,\cN}^{exp}] \sim \xi_L\left(1-\frac{1-y}{\cN C_L g(\alpha)}\xi_L^{\tau-1}\right)
\eeq

Since $\tau >1$ the correction term increases as $L$ grows.
It eventually becomes of the same order as $\xi_L$, but when $L$ increases
one has to add higher order corrections to eq. (\ref{approxxi}).
 On the other hand, were $\tau <1$, then
would the correction term become negligible as $L$ grows.
   
The $L$ value where the effect starts can be
estimated by :

\beq\label{Lmax}
L \sim \left(\frac{1-y}{\cN C_L g(\alpha)}\right)^{\frac{-1}{\beta(\tau-1)}}
\eeq

Consequently, this effect is more prominent when $\beta(\tau-1)$ is larger.\\

The ratio  $\alpha_L=\frac{E[\xi_{L,\cN}^{exp}]}{L^\beta}$ is therefore not equal
to a constant $\alpha$ as it must be, 
but is $L$ dependent (see Fig. (\ref{Powcutoffdeg} a)).
Clearly, for sufficiently large $L$
\textit{the corresponding probability violates the finite size scaling and 
 the data collapse}. Furthermore, the scaling of the moments 
is also affected by this effect.
Indeed the moments are obtained empirically from the formula
$m^{exp}_L(q)=\sum_{n=1}^{\xi_L^{exp}}n^qP_L^{exp}(n)$
and the scaling exponents are  extrapolated
from the formula:
$$
\sigma^{exp}(q)= \lim_{L\to \infty} \frac{\log(m^{exp}_L(q))}{\log(L)}
$$
When $L$ is sufficiently small, $\xi^{exp}_L \sim \alpha L^\beta$
since the correction due to the finite
sampling has essentially no effect. Then one obtains the right
scaling exponent $\sigma(q)$ from the data. However, when $L$ increases, 
one observes
\textit{ a spurious deviation of the curve $m_L(q)$ from the theoretical value}. 
This effect is illustrated
Fig. \ref{Powcutoffdeg} in the case $\tau=1.9,\beta=2,\gamma=2.5,\alpha=2$,
$g(x)=e^{-x^\gamma}$ where two samples with $\cN=10^6$ and
$\cN=10^8$ were generated. Note that the effect is more prominent
when $q$ increases.

\bef  
\bc
\begin{minipage}{6cm}   
\epsfxsize=6cm
\epsfysize=6cm  
\epsffile{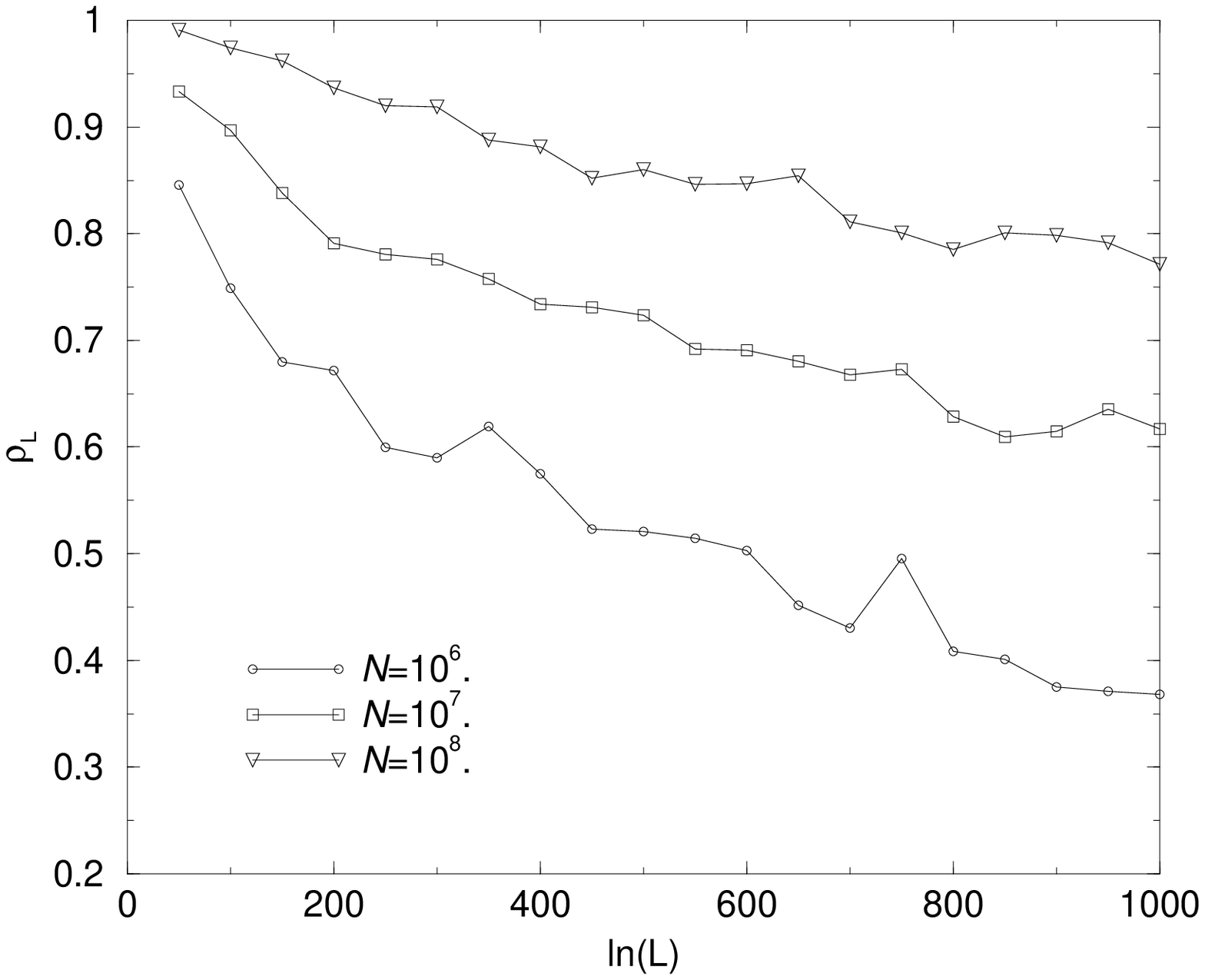}
\end{minipage} \hspace{1cm}
\begin{minipage}{6cm}
\epsfxsize=6cm  
\epsfysize=6cm
\epsffile{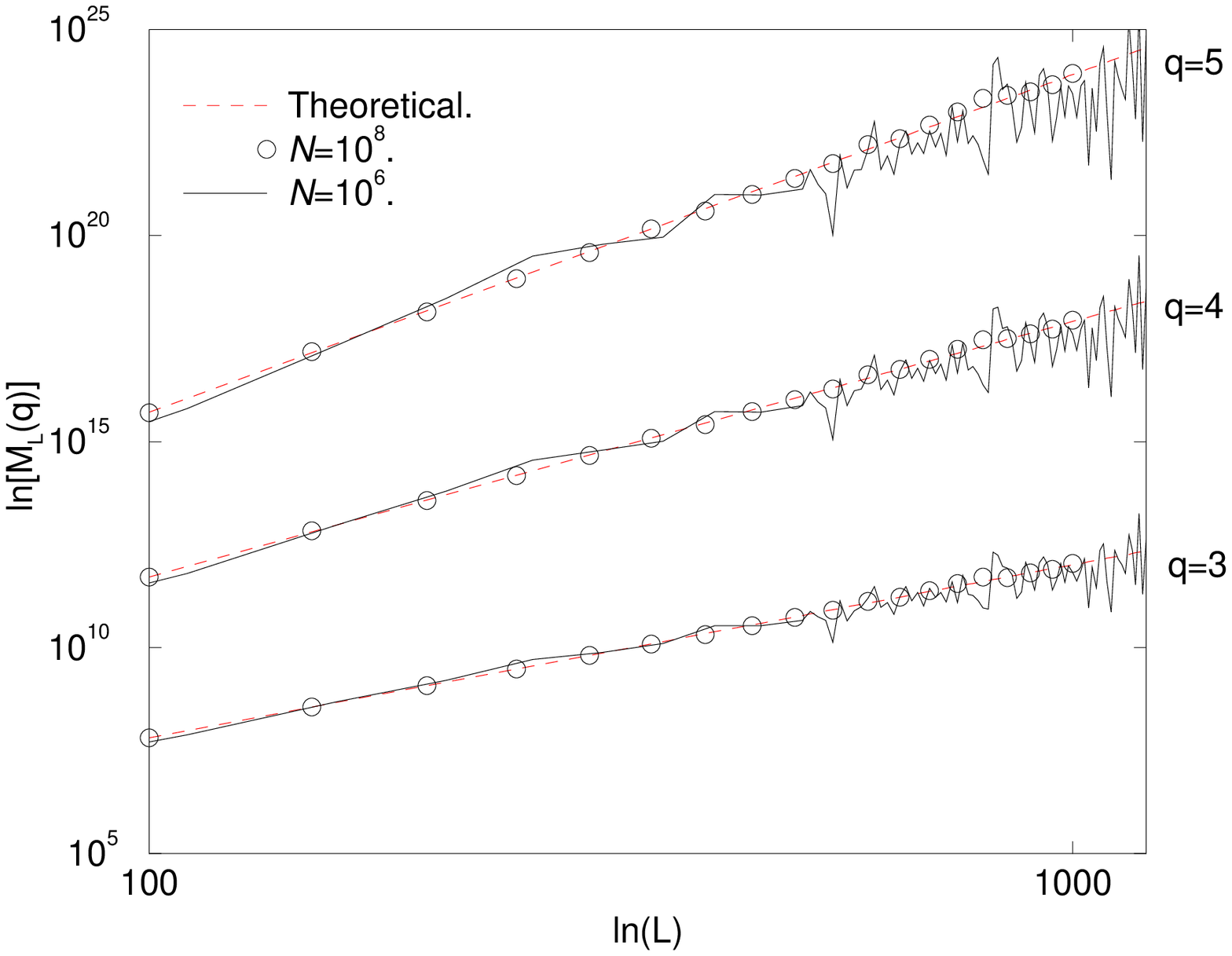} 
\end{minipage}
\vspace{0.3cm} 
\caption{Effect of the sampling cut off.
Fig. \ref{Powcutoffdeg} a) Ratio  
$\rho_L=\frac{E[\xi_{L,\cN}^{exp}]}{\xi_L}$. Fig. \ref{Powcutoffdeg} b) Moments.}
\label{Powcutoffdeg}
\ec
\enf

This computation shows therefore that for the values $\tau >1$ 
numerical problems
appears, induced by the finite size sampling.
Not only the fluctuations of $\xi_{L,\cN}^{exp}$  increase
but also the averaged value $E[\xi_{L,\cN}^{exp}]$ is biased. To our
opinion, the estimation of the correct $\xi_L$ is the main problem in
analyzing the data from SOC simulations. 

The analysis of the Lee-Yang zeroes for relatively small sizes can
however give a fairly good estimate of the values $\alpha,\beta$ allowing
to extrapolate $\xi_L$ to larger size. Indeed,  eq. (\ref{tzeroesFSS}) suggests that
the argument of $t_L(k)$ is not too sensitive to the fluctuations of
$\xi_{L,\cN}^{exp}$ (compared to the fluctuations of the moments
which are of order $\left(\xi_{L,\cN}^{exp}\right)^{q+1-\tau}$).
Hence, it is possible to find the values of $\alpha,\beta$. 
We give an example Fig.\ref{alphaKada} where the empirical data are the same as those
used for the computation of the moments. As for the truncated
power law, we found a slight deviation for
the first zero. Interpolating the values from $k=1 \dots 10$ we found
$\alpha= 2.094 \pm 0.006 \pm 0.09, \beta=
2.002 \pm 0.003$ which gives quite a good estimate.

\bef  
\bc
\begin{minipage}{6cm}   
\epsfxsize=6cm
\epsfysize=6cm  
\epsffile{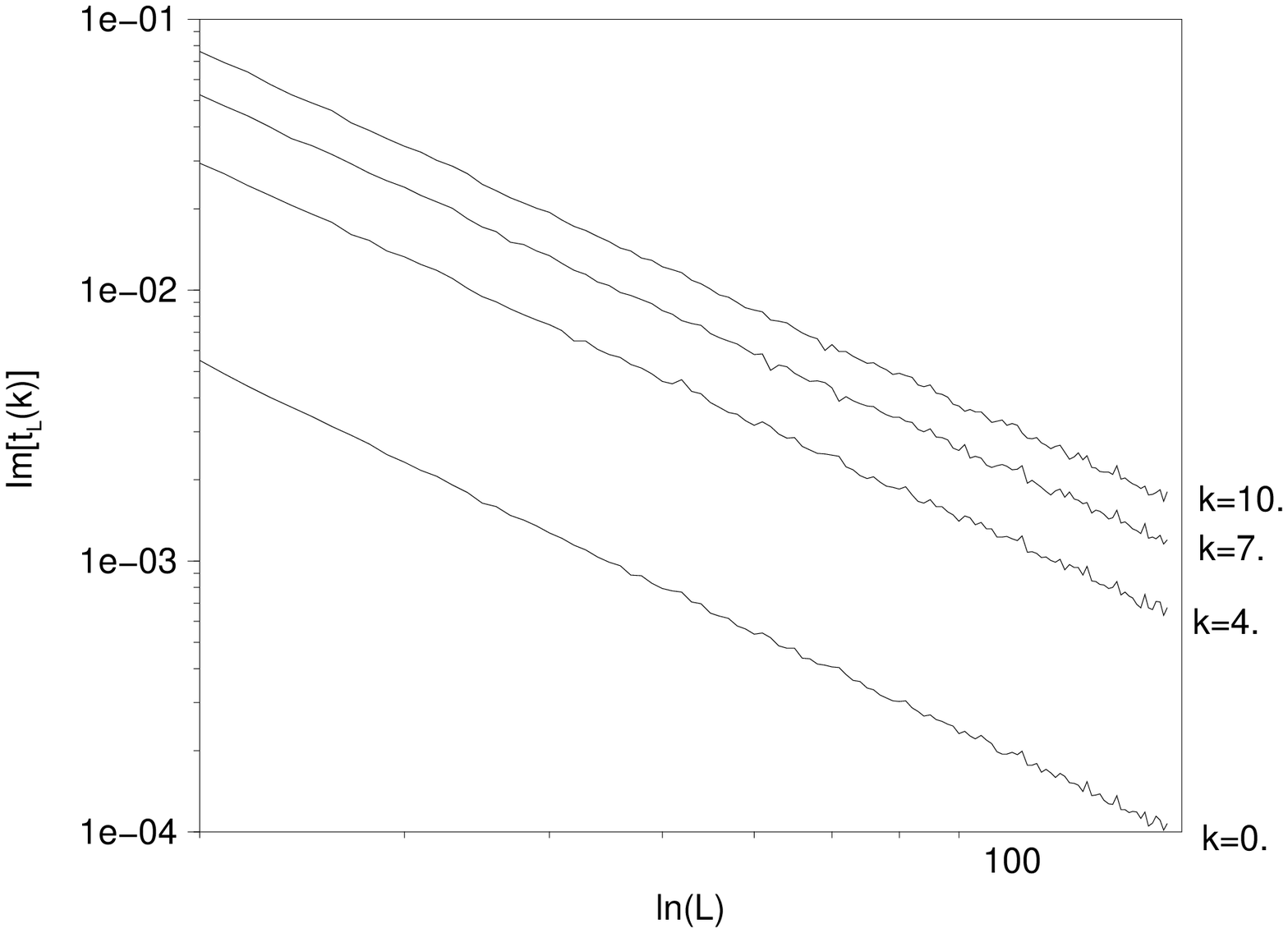}
\end{minipage} \hspace{1cm}
\begin{minipage}{6cm}
\epsfxsize=6cm  
\epsfysize=6cm
\epsffile{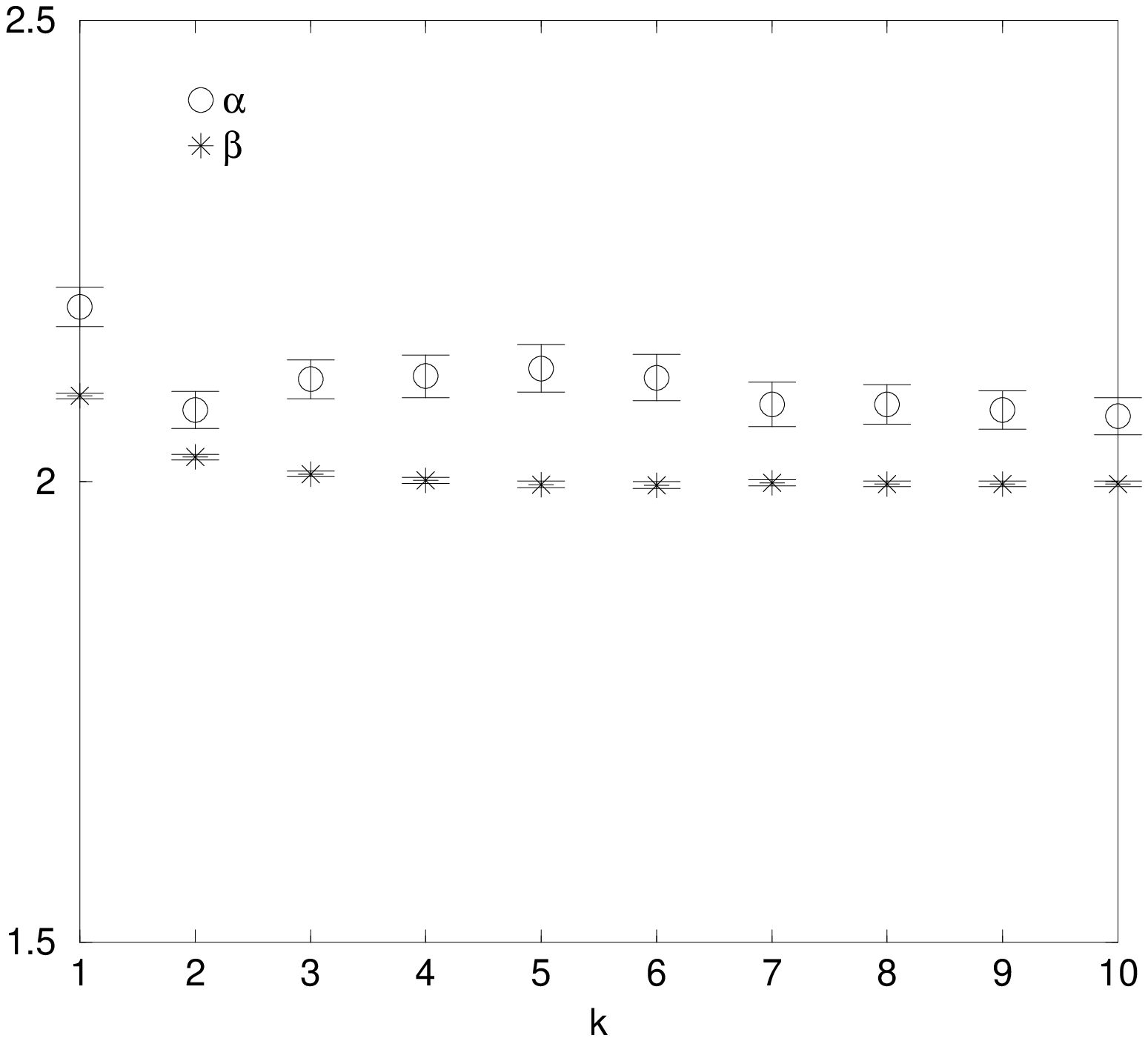} 
\end{minipage}
\vspace{0.3cm} 
\caption{ a) Argument of the first Lee-Yang zeroes
of an empirical distribution generated from $\cN=10^8$ samples.
b) Values of $\alpha,\beta$ extrapolated from a).}
\label{alphaKada}
\ec
\enf

The real part of Lee-Yang zeroes and consequently, the argument,
is more  sensitive to non extensive sampling  effect. 
An analytical expression is obtained if one modifies
 the equations (\ref{tzeroespowR},\ref{tzeroespowangle},\ref{Argpow})
 by replacing $\alpha$  by $\alpha_L$. The argument of $t_L(k)$ writes now:

\beq
Arg(t_L(k))\sim
\arctan\left(
\frac
{2k\pi}
{\left[-\log(\frac{\alpha_L\Gamma_g(\tau-1)}{\Lambda_L^{\tau-1}}) +
\frac{1}{2}\log\left( \log^2(\frac{\alpha_L\Gamma_g(\tau-1)}
{\Lambda_L^{\tau-1}})
+4k^2\pi^2\right)
 \right]}
\right)
\eeq

The finite sample effect is illustrated Fig. \ref{Powcutoffarg} for the first zero.
One observes a deviation from the real curve for $\cN=10^6$. 
This can be used as an empirical
way to define the $L$ where the empirical distribution is not biased.
Note that the curve of the argument of the first empirical zero 
obtained for $\cN=10^8$ follows the theoretical curve with a good accuracy.
This shows that the determination of the zeroes is robust with respect to
fluctuations in the coefficients of the polynom (\ref{ZL}). This is in fact
an easy consequence of the implicit function theorem.
  
\bef
\bc
\begin{minipage}{6cm}
\epsfxsize=6cm
\epsfysize=6cm
\epsffile{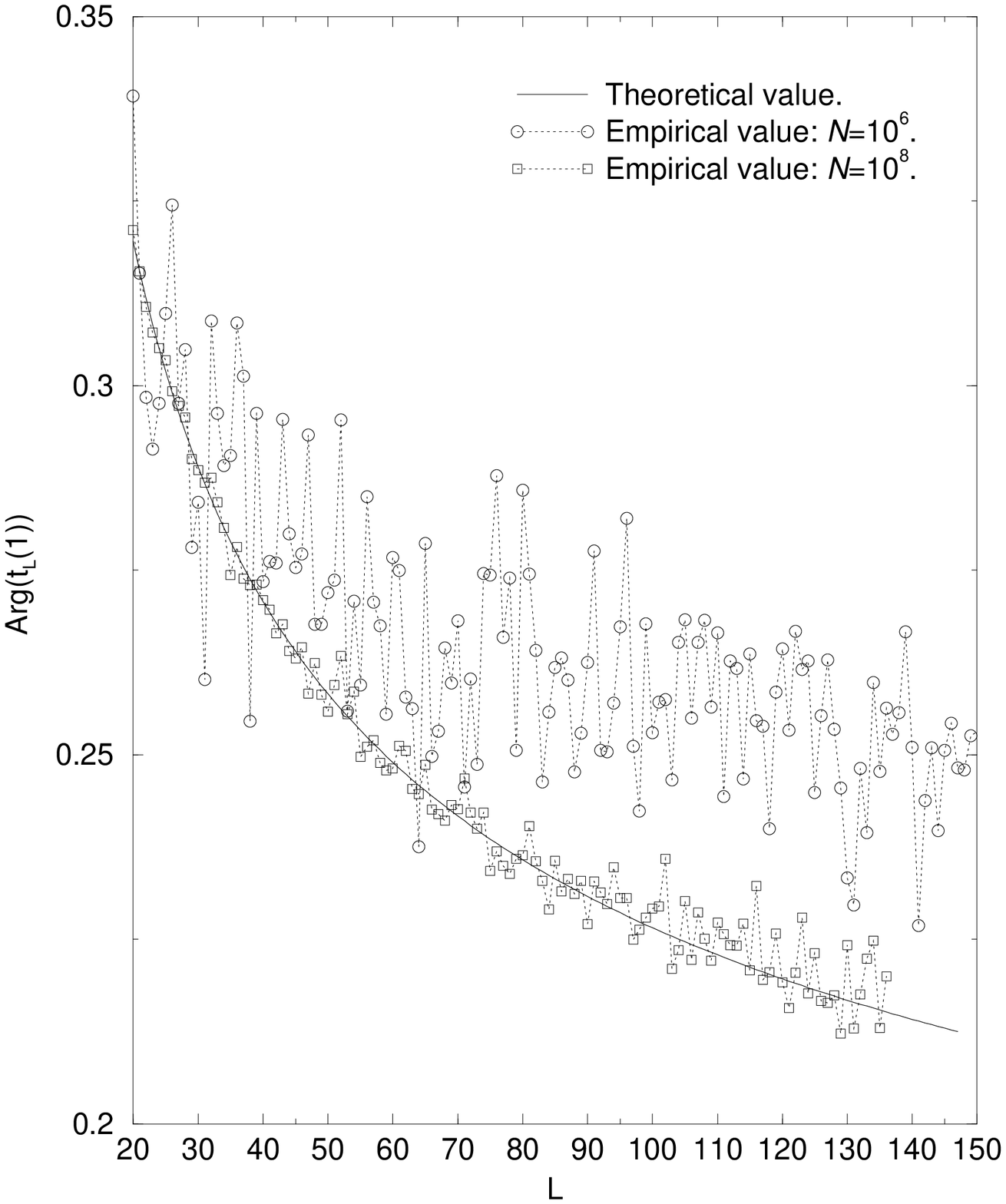}
\end{minipage}
\vspace{0.3cm}
\caption{ Effect of a sample cut off
on the argument of the zeroes in the $t$ plane.}
\label{Powcutoffarg}
\ec
\enf

Looking at fig. (\ref{Powcutoffdeg}b) and fig. (\ref{Powcutoffarg}) one could 
argue that the moments are less
sensitive than the Lee-Yang zeroes to a size independent sampling. However,
the sensitivity of the moments increases with the degree $q$. Therefore,
to detect this effect, one has to compute the moments for high $q$. This
can clearly cause
numerical problems. On the other hand, the Lee-Yang zeroes
integrate informations about all order moments and the sensitivity
can be detected easily. We believe that this kind of analysis is prior
to any investigation concerning in particular the multifractal nature of the scaling.   

Our conclusion  is therefore twofold. Firstly, some cautions are required
when  increasing $L$ to extrapolate the thermodynamic limit.
If $\cN$ is kept fixed, \textit{too large} $L$ will give \textit{wrong} estimations.
In this case at least, the bigger  is  not the best.
This effect can be compared to the critical slowing down in the literature about
critical phenomena. However, to the best of our knowledge we don't know
any example in the SOC literature where this effect has been discussed.
Secondly,
the Lee-Yang zeroes give nevertheless useful informations. They can be used
to determine the range of $L$ values where the data are not too affected,
and, in this range, the simple scaling of the imaginary part allows to determine
the scaling of $\xi_L$. This can used with other methods such as binning, or smoothing
to determine the exact distribution from the empirical one.

\sssu{Other scaling.} \label{PAC}

In this section we investigate briefly the other scaling forms discussed in section
\ref{CO}. Our main conclusion is that the Lee-Yang zeroes are
highly
 sensitive to
the changes in the scaling form. This remark opens perspective
to develop a general theory allowing to extrapolate the characteristic
of probability distributions from the behavior of the Lee-Yang zeroes
of the empirical generating function. 
However, we have not yet been able to 
provide  an
equivalent of the analytic forms (\ref{SolPow},\ref{SolKada}) that would be 
helpful to properly extract the features of the probability distribution
from the Lee-Yang zeroes. The development of such a general framework
 is under investigation and will be published
in a separated paper. We give a few numerical
 examples fig. \ref{FPac}a, b.

We investigated first the case (\ref{PPac})
where  $F_L(x)=Y(x-\beta_L)$. In this case, the computations
done in section \ref{POW} essentially hold, with a $\beta$
depending  on $L$. In particular, eq. (\ref{zzeroespowtheta}) suggests that
the $L$ dependence of $\beta$ should be detected on the argument
of $z_L(k)$, $\theta_L(k)$. In fig. \ref{FPac}, we plotted $\theta_L(5)$
in the case $\beta_L=\beta(1-\frac{1}{L})$, where $\beta=2,
\tau=1.9$, and $L=20 \dots 120$. The theoretical prediction is $\theta_L(5)=
\frac{10\pi}{L^{\beta(1-\frac{1}{L})}}$. We tried a fit
of the form $f(x)=\frac{10\pi}{L^{\beta(1-\frac{1}{L^\alpha})}}$,
where $\alpha,\beta$ are the fit parameters. We found $
\alpha=1.014(8) \pm 0.0006, \beta=2.013(9)\pm 0.0001$,
which is quite satisfactory.

We also tried a more general form for $F_L(x)$
with a non linear $F_L$ converging to a step function
as $\L \to \infty$:

\beq \label{ExFL}
F_L(x)=-(1+\tanh\left(L^{\alpha_1}\left(x-\beta-\frac{1}{L^{\alpha_2}}\right) \right));
\qquad x \leq \beta-\frac{1}{L^{\alpha_2}}
\eeq

In our simulations, $\beta=2,\tau=1.9,\alpha_1=0.1,\alpha_2=1$.
$L^{\alpha_1}$ controls the rate of approach of $F_L(x)$ to the step function as $L$
grows. A small $\alpha_1$ gives a slow convergence and therefore
has an effect up to very large $L$.

The result for the argument $\theta_L(5)$ is also represented fig. \ref{FPac}a.
We note that the curve is indistinguishable from the previous case
and therefore the imaginary part of $t_L$ is not sensitive to the non linear effect
of the $\tanh$,
but gives the right $\alpha_2$. On the other hand, we noted that the
argument of  $t_L$ is sensitive to the non linear effect
(fig. \ref{FPac}b). \\

For the multifractal case we studied the case where the multifractal
spectrum has the form $f(x)=C-\tau x
-a x^2$. This is the lowest degree non linear form of $f$ compatible
with (i),(ii) and with the convexity  of $f$. The values
of $\alpha=1,\beta=2,\tau=1.9$ were the same as for the previous examples.
We observe (Fig. \ref{FPac}a) that $\theta_L(k)$ is not sensitive to
the multifractality and gives therefore the right $\alpha,\beta$.
Consequently, our method to estimate the degree is also valid
for a multifractal distribution (\ref{KadaMF}). On the other hand $Arg(t_L(k))$
is modified for a multifractal distribution, but we are not yet able
to analyze this variation.

All the results are depicted fig. \ref{FPac}a ($\theta_L(5)=Im(t_L(5))$)
and fig. \ref{FPac}b ($Arg(t_L(5))$).

\bef  
\bc
\begin{minipage}{6cm}
\epsfxsize=6cm
\epsfysize=6cm
\epsffile{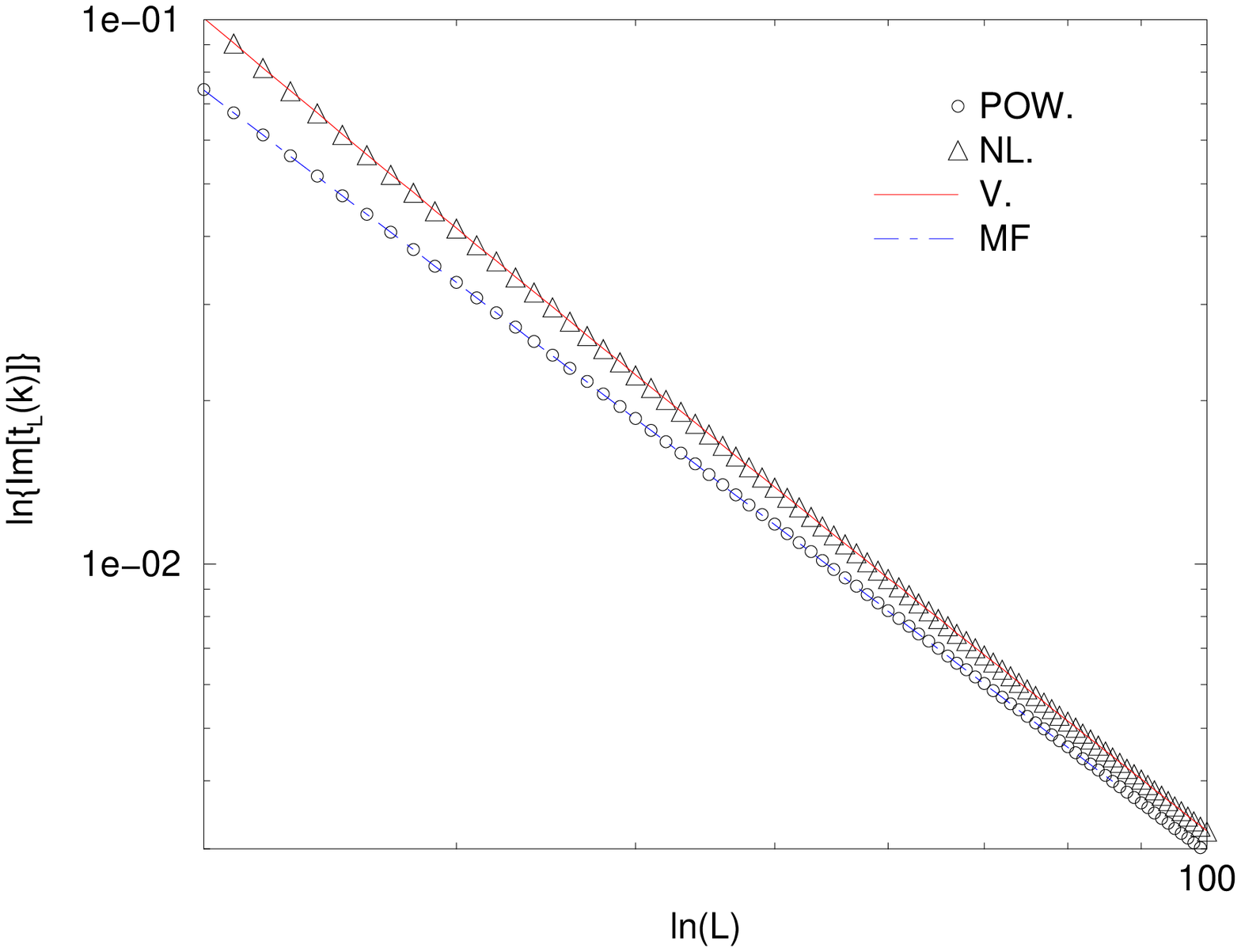} 
\end{minipage} \hspace{1cm}
\begin{minipage}{6cm}
\epsfxsize=6cm
\epsfysize=6cm
\epsffile{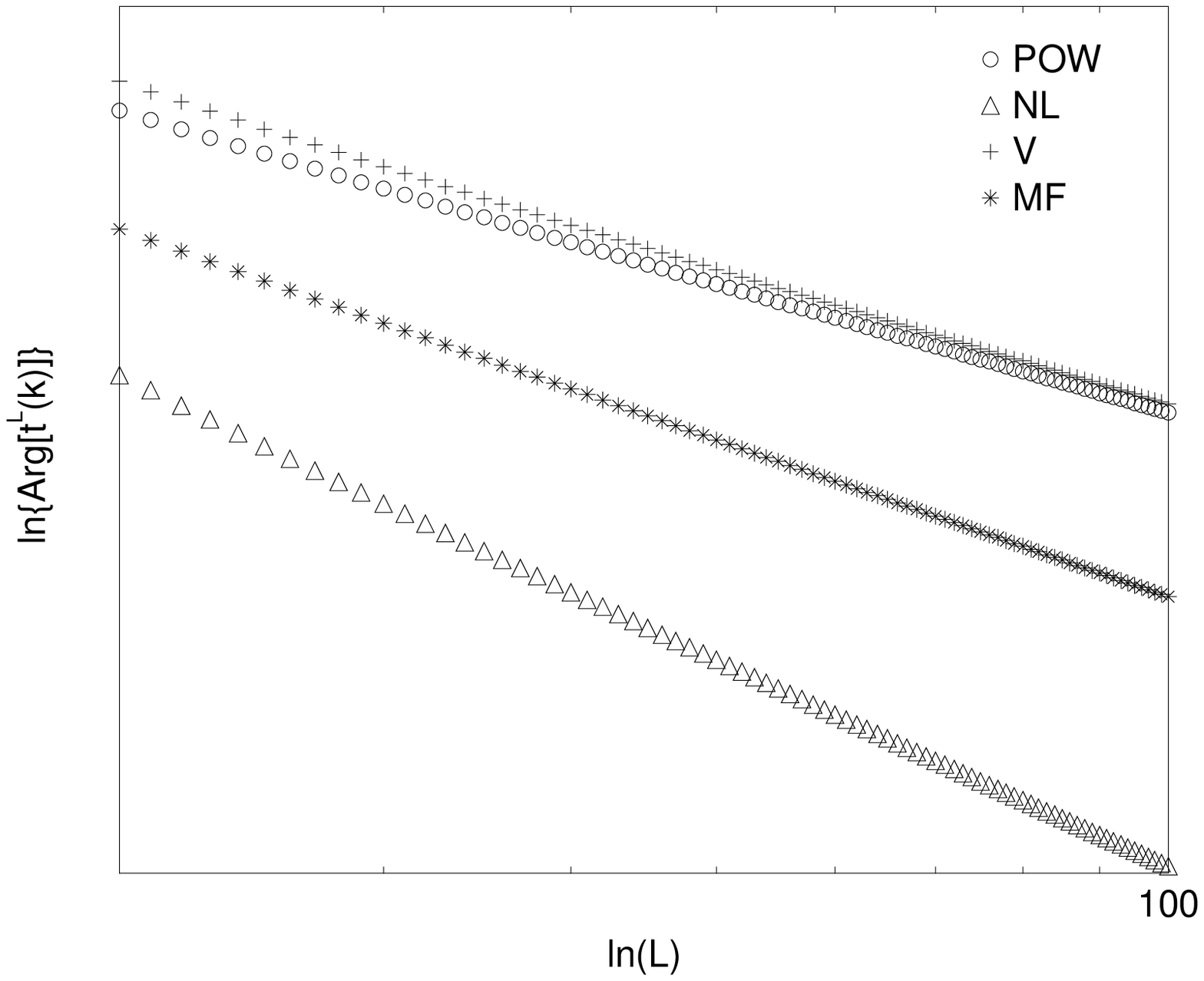}
\end{minipage} 
\vspace{0.3cm}
\caption{Fig. \ref{FPac}a. Argument $\theta_L(5)$ for the truncated
power law (POW); a truncated power law with $L$ dependent $\beta$
(V);  a probability distribution of the form (\ref{PPac})
with $F_L(x)$ given by (\ref{ExFL}) (NL) and
a multifractal distribution (MF).Fig. \ref{FPac}b. Argument
of $t_L(5)$.}
\label{FPac}
\ec
\enf

\su{An example of SOC model.} \label{ZHANG}

In this section, we study an example of  
 Lee-Yang zeroes computation for a SOC model, the Zhang
model, defined as follows.

Let $\Lambda$ be a d-dimensional sub-lattice in $\bbbz^d$, taken as a
square of edge length $L$ for simplicity. Call $N=\#\Lambda=L^d$, and
let $\dL$ be the boundary of $\Lambda$, namely the set of points in
$\bbbz^d \setminus\Lambda$ at distance $1$ from $\Lambda$.  Each site
$i \in \Lambda$ is characterized by its "energy" $X_i$, which is a
non-negative real number. Call  $\bX = \lbrace X_i
\rbrace_{_{i \in \Lambda}}$ a configuration of energies.
Let $E_c$ be a real, strictly positive number, called the {\it critical energy},
and
$\cm = [0,E_c[^N$. A configuration
$\bX$ is called "stable" if $\bX \in \cm$
and "unstable" otherwise. If $\bX$ is stable then
one chooses a site $i$ at random with  probability $\frac{1}{N}$, and add to
 it energy
$\delta$(excitation).
If a site $i$ is \textit{over critical} or \textit{active}
 ($X_i \geq E_c$), it loses a part of its energy
 in equal parts to its $2d$ neighbors (relaxation).
 Namely, we fix a parameter
$\epsilon \in [0,1[$ such that
the remaining energy of $i$ is $\epsilon X_i$, after relaxation of the site $i$,
 while the $2d$ neighbors
receive the energy $\frac{(1-\epsilon)X_i}{2d}$. Note therefore that the  
energy is locally conserved.
If several nodes are simultaneously active,
the local distribution rules are
additively superposed,
i.e. the time evolution of the system is synchronous.  The succession
of updating leading an unstable configuration
to a stable one is called an {\it avalanche}.
 The energy   is dissipated at the boundaries of the system,
namely
the sites of $\dL$ have always
zero energy. As a result, all avalanches are {\it finite}.
Consequently, whatever the observable $n$, $\xi_L < \infty$
for finite $L$.
The addition of energy is {\it adiabatic}. When an avalanche occurs,
one waits until it stops before adding a new energy quantum.
Further excitations eventually generate a new avalanche,
but, because of the adiabatic rule, each new avalanche starts from {\it only one}
active site.
It is conjectured that a critical state is reached in the thermodynamic limit.

 Though it has long been believed that the Zhang model obeys
finite size scaling (\ref{Kada}), a recent paper revised this point
of view and claimed that the Zhang model
 does not even have a multifractal scaling
(but no alternative scaling
was proposed \cite{Vespignani1}). We will not solve this debate in this paper. 
Rather
we  will come to two  conclusions. Firstly,
 because of high sensitivity of the model to
the sample cut off (fig. \ref{angle_t}),  
one has somehow to relativise the conclusions about the scaling
obtained from the numerical simulations.
This shows that to draw any reliable conclusion on the scaling
one has to increase \textit{the sample with the system size (e.g. like $L^\beta$)}.
This will clearly be rapidly intractable even for the fastest computers. 
Secondly, the Lee-Yang zeroes gives rather reliable extrapolations provided
the size $L$ is not too large.\\

We computed the empirical probability distribution of avalanche sizes
$P^{exp}_L(s)$ where the size is the total number of relaxing
sites during one avalanche. We did our simulations
for lattice sizes from $L=10$ to $L=55$
in two dimensions, with $E_c=2.2,\epsilon=0.1$ with
a statistics over $\cN = 10^6$ and $\cN=10^8$ avalanches.
Consequently, $\cN$ was fixed \textit{independently of $L$}
as usually done in SOC numerical simulations.

We first present Fig. \ref{Plinterpol}a
the experimental zeroes in the $z$ complex plane for $L=30,\ \cN=10^6$.
 In order to see the effect of the noise, we also computed
the zeroes of a smoothed version of the empirical probability distribution
(fig. (\ref{Plinterpol}b)). The smoothing method uses a binning procedure,
followed by a spline extrapolation, allowing to fill the ``holes''
existing in the empirical probability distribution. These holes corresponds
to events that didn't happen during the trial and consequently are
given a zero probability. In the numerical computation of the zeroes,
these holes corresponds therefore to vanishing coefficients in the polynom
(\ref{ZL}), that produce problems in the convergence of most of the root
finding algorithms. Our numerical procedure seems however to be robust
with respect to this effect. 

\bef
\bc  
  
\begin{minipage}{6cm}
\epsfxsize=6cm  
\epsfysize=6cm
\epsffile{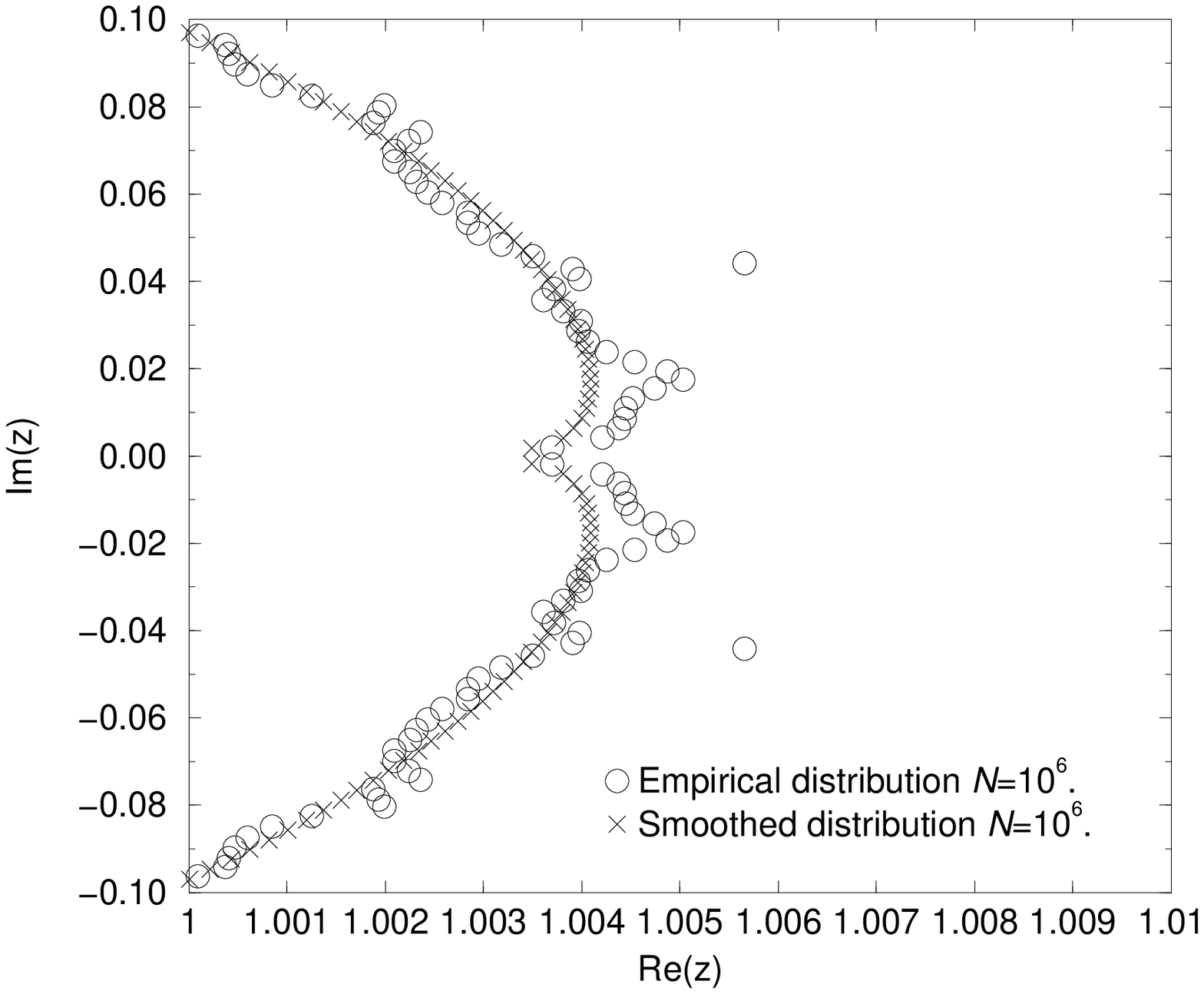} 
\end{minipage}
\begin{minipage}{6cm}
\epsfxsize=6cm  
\epsfysize=6cm
\epsffile{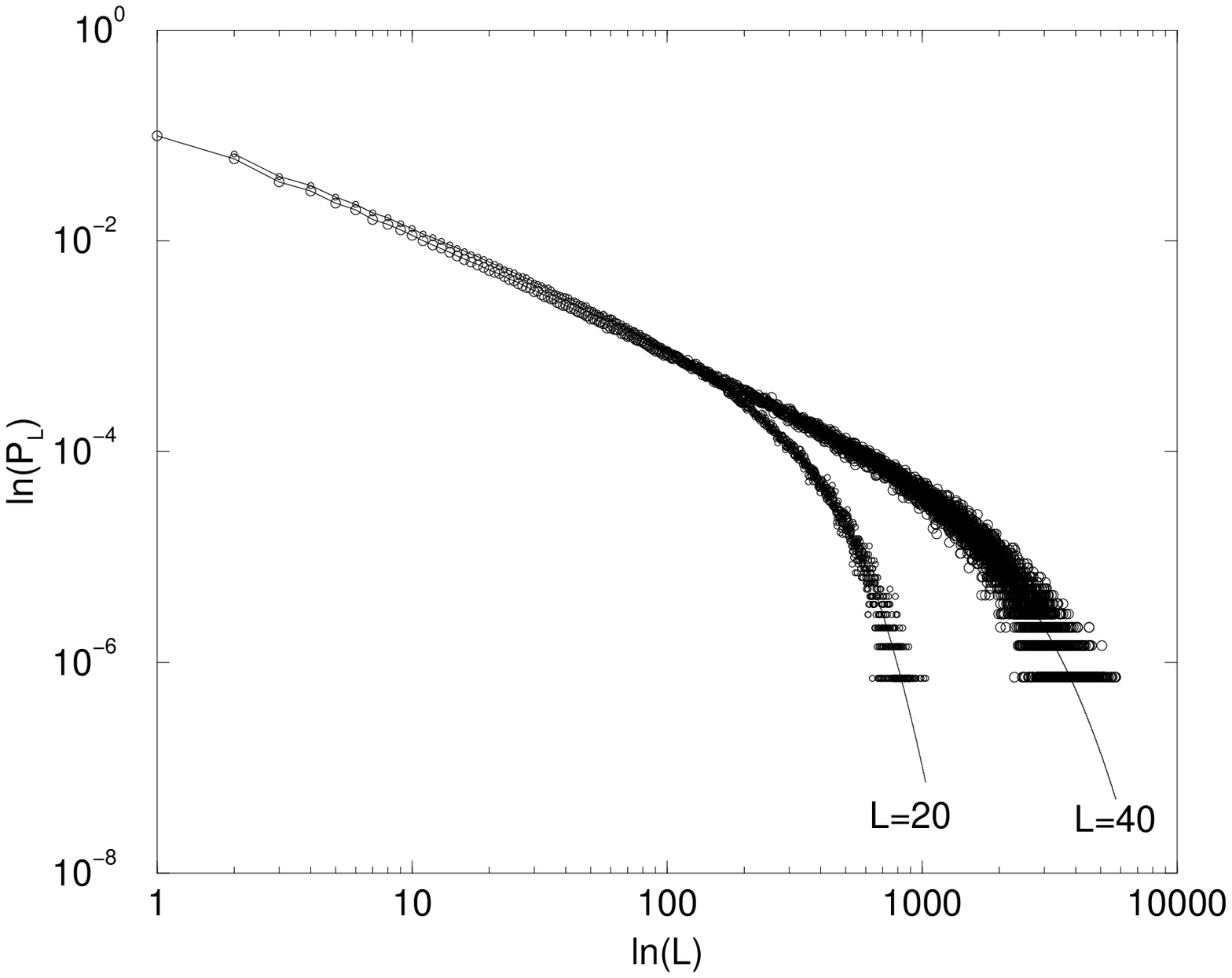} 
\end{minipage} \hspace{1cm}
\vspace{0.3cm}  
\caption{a) Zeroes in the $z$ plane for $L=30$,
$\epsilon=0.1$, $\cN=10^6$, $E_c=2.2$..
The zeroes of the
experimental, $\cN=10^6$, and smoothed probability distribution
are represented. b) Examples of empirical and smoothed probability
distributions used in the computation of the zeroes
}
\label{Plinterpol}
\ec
\enf           

 As argued all along in this paper, the argument
$\theta_L(k)$ provides a way to determine the exponent $\beta$
characterizing the maximal avalanche size. We plot fig. \ref{argument_z_Zhang}a
the $\theta_L(k)$'s
for $k=1 \dots 5$ versus $L$. We note that $\theta_L(k)$ is quite
robust to noise and gives therefore a reliable way to
measure $\beta,\alpha$.
We used a fit form $\frac{2\pi k}{\alpha L^\beta}$, where
$\alpha, \beta$ are fit parameters. We found a slight $k$
dependence for the first zeroes (as expected from eq. (\ref{tzeroespowangle}) if
one assumes FSS). We plotted fig. \ref{argument_z_Zhang}b
the extrapolated $\alpha, \beta$ versus $k$. For $k >3$
these value seem to stabilize around $\alpha=0.62 \pm 0.07$
and $\beta=2.59 \pm0.04$. In the finite size scaling ansatz 
$\beta$ and $\tau$ are related by $\beta(2-\tau)=2$ \cite{BCK4} and
$\tau=1.253$ is known from the renormalization group analysis.
Therefore the predicted value for $\beta$ is $2.667$.
Despite the smallness of the $L$ we considered, the computed value
is not too far from the predicted one. However, an
 accurate determination of
$\beta, \tau$ and a precise check of FSS would demand somewhat larger size systems,
that we were unable to generate for this illustration. (Note that the main
problem is not the computation of the zeroes since
there exist quite fast and precise root finding algorithms,
 but the generation of $P_L^{exp}$ itself.).

\bef
\bc
\begin{minipage}{6cm}
\epsfxsize=6cm  
\epsfysize=6cm
\epsffile{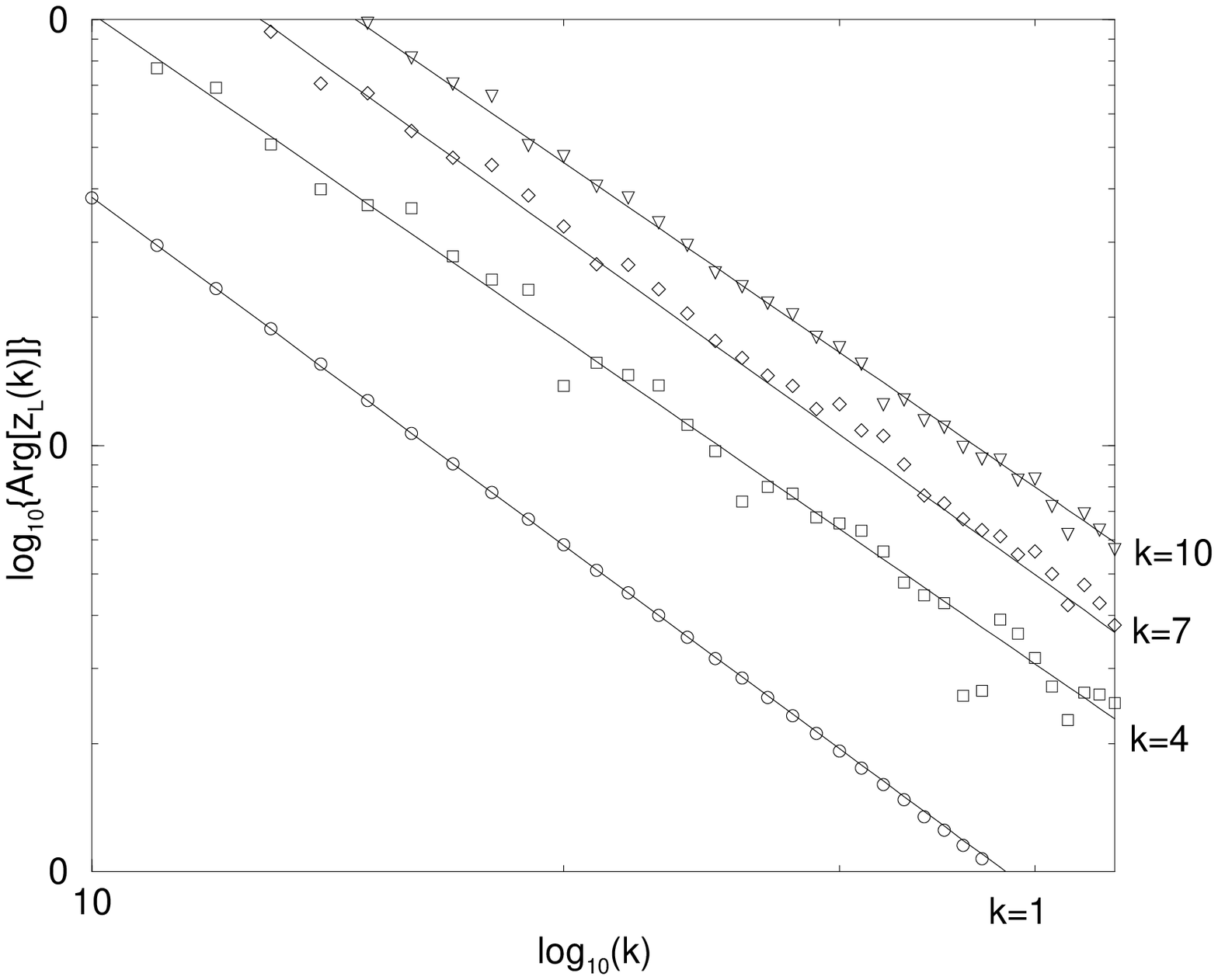} 
\end{minipage} \hspace{1cm}
\begin{minipage}{6cm}
\epsfxsize=6cm
\epsfysize=6cm
\epsffile{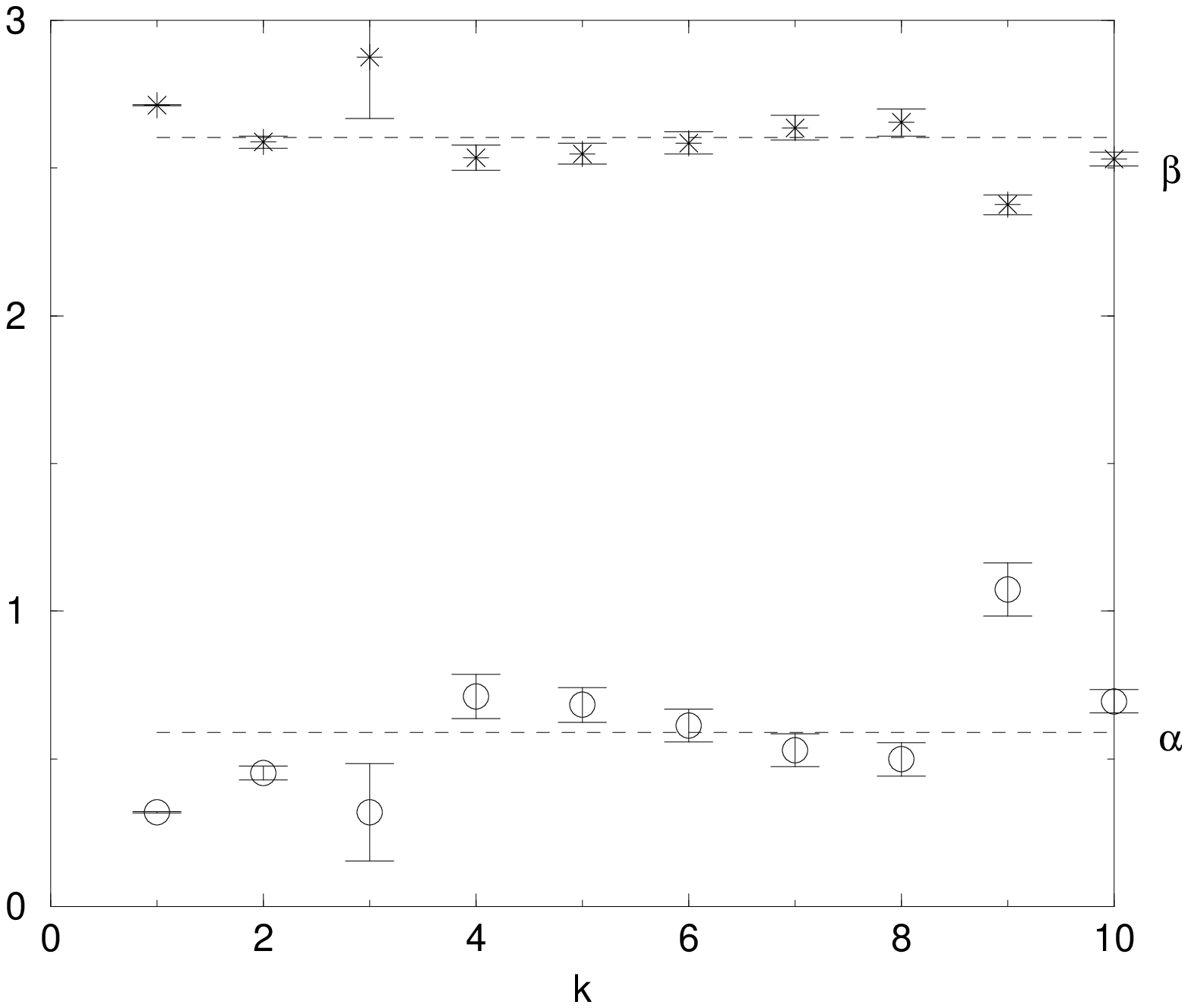}
\end{minipage}
\vspace{0.3cm}
\caption{a. Argument of the 5 first Lee-Yang zero in the z plane
versus $L$, for  $Ec=2.2,\epsilon=0.1$.
The full lines corresponds to a fit. b. Plot of the fit parameters $\alpha,\beta$
versus $k$ .
\label{argument_z_Zhang}}
\ec
\enf

Finally, we investigate the scaling of the argument $t_L(k)$
and a possible  size independent sampling effect. The main difficulty here
are the wild fluctuations in $\xi_L^{exp}$. Indeed, the real part
of $t_L(k)$ is more sensitive than the imaginary part to these
fluctuations and consequently $Arg(t_L(k))$ as wild fluctuations.
 Only the first zeroes
seem to be robust to this effect.  As discussed in section
\ref{cutoff} these fluctuations in $\xi_L^{exp}$ are intrinsic to
the empirical computation of $P_L^{exp(L)}$ and cause problems
in the extrapolation to the thermodynamics limit, whatever the method
used. We plotted Fig.\ref{angle_t} the argument of the 2 first Lee-Yang zeroes
 in the $t$ plane, versus $L$. Our simulations suggests that  the Zhang model 
is sensitive to the size independent sampling.
Note in particular that the values of $L$ that we used in our computation are
quite smaller that the ones found in the literature and the effect is already
significant for $\cN \leq 10^6$. 

\bef
\bc
\begin{minipage}{6cm}
\epsfxsize=6cm
\epsfysize=6cm
\epsffile{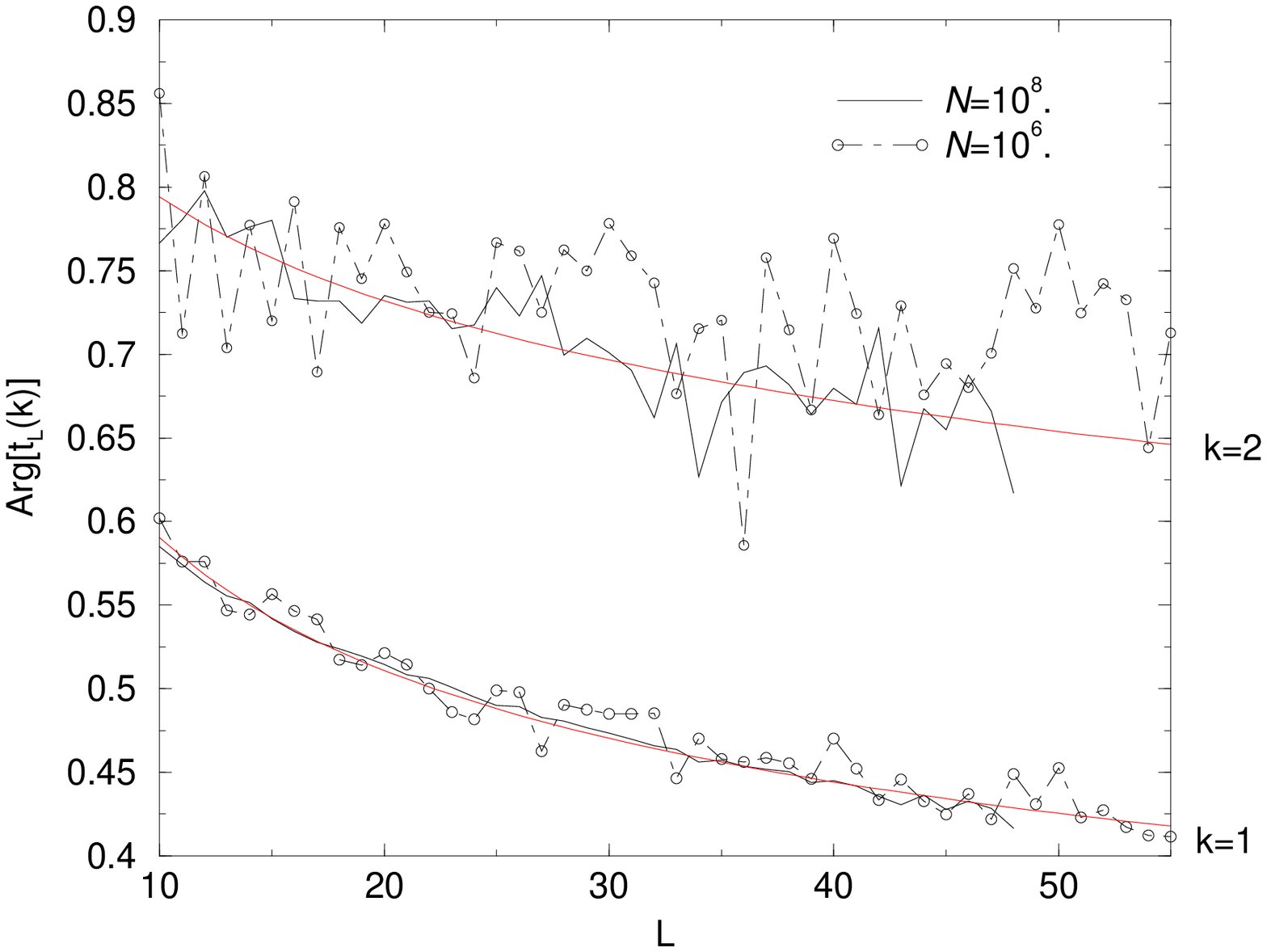}
\end{minipage}
\vspace{0.3cm}
\caption{Argument of  Lee-Yang zeroes $t_L(1), \ t_L(1)$
for  $Ec=2.2,\epsilon=0.1$. In color full lines are drawn the interpolation
of the empirical curves obtained from the sampling with $\cN=10^8$. 
\label{angle_t}}
\ec
\enf
  
This shows clearly that
a  re estimation of the conclusions drawn from the numerics in the
Zhang model (and also, maybe, for some other SOC models)
 should be done in light of this effect.
On the other hand, our approach suggests that there may be no need
to go to gigantic sizes provided the finite lattice size effects are carefully
handled. From this point of view, analytic formula like
(\ref{argtFSS})  might provide a way to analyze these effects.

\su{Conclusion.}

In this paper, we have shown that the finite size study of the SOC like 
probability distributions
leads to similar Lee Yang or Fisher phenomenon as in statistical physics models
near the critical point. This implies that the convergence of the SOC state to a critical
state with power law statistics can be analyzed in a similar way as equilibrium
statistical mechanics.
More precisely, the way the zeros of the partition function accumulate on 
the real axis, when
the size of the system grows up, provides relevant 
informations on the critical structure of the
observed system. In particular, it permits to measure useful critical indexes of the
underlying theory.

 Moreover, we have shown that the size of the SOC models power exponent,
$\tau>1$, leads to a comprehensive violation of the standard scaling laws.
We give a approximate theory of this effect well confirmed by numerical simulations.
It is the same characteristic which leads to a specific sensibility  of the SOC
numerical experiments to size independent sampling effects. 
We studied carefully this
 effect on extrapolation to the limit $L\to \infty$ and show that it could possibly
mimics important effects such as multifractality. We notice that the argument
of the zeros in the $t=log(z)$ plane is a good test of this effect.

On one other hand, we show that the arguments of the first zeros in the $z$ plane
of the generating function, $G(z)$ is rather insensitive to these effects,
statistically robust, and provide a nice way to compute the SOC $\alpha$ and
$\beta$ parameters. Using the standard Kadanoff et al. scaling form  \cite{Kadanoff},
 we verify that the parameter's values as extracted
from numerical simulations where in good agreement 
with the theoretical input of the model.
This last result gives us some confidence to extract the values of these parameters
from Zhang's model  numerical data. 
Notice that these results have been extracted from medium range simulations. 
This shows up
once more the power of the finite  size analysis of the critical phenomenon.

This paper is (with \cite{BCK5}) a first step toward a scaling
theory of SOC system from the behavior of the Lee-Yang zeroes. The next
step would be the definition of the exponents characterizing the approach
to criticality, like the exponents $\alpha,\beta,\gamma$ in statistical mechanics
and their link to the scaling of the zeroes.\\

{\bf Acknowledgments.}
This work has been partially supported by the Zentrum fuer Interdisciplinaere
Forschung (ZIF) of Bielefeld (Germany), in the frame of the
projet
"The Sciences of Complexity: From Mathematics to Technology to a Sustainable World".
B.C. warmly acknowledge the ZIF for its hospitality.
 He also thanks the CNRS
for its support allowing him to have one free year without teaching.
He is grateful to Ph. Blanchard, T. Krueger, P. Bak, and M. Paczuski for
illuminating discussions in Bielefeld. We also thank G. Batrouni and K. Bernardet
for computer facilities and useful references.

\ed

\bibitem{Pietronero} Pietronero L.,Vespignani A., Zapperi S., Phys. Rev. Lett. 72;
1690 (1994); Vespignani A., Zapperi S.,Pietronero L., Phys. Rev. E., 51:1711 (1995).

\bibitem{Vespignani2} Vespignani A., Zapperi S.,
``How self-organized criticality works :
A unified mean-field picture.''
{\em Physical Review E57, 6345} (1998)